\documentclass[11pt]{article}

\usepackage[leqno]{amsmath}
\usepackage{amssymb}
\usepackage{enumerate}
\numberwithin{equation}{section}

\allowdisplaybreaks

\usepackage{graphicx}
\usepackage{color}

\newcommand{\p}{\partial}
\newcommand{\e}{\epsilon}
\newcommand{\lm}{\Lambda}

\newcommand{\tp}{\tau_{\mathrm{top}}}
\newcommand{\up}{u_{\mathrm{top}}}
\newcommand{\vp}{v_{\mathrm{top}}}
\newcommand{\vh}{v_{\textsc{\tiny\rm H}}}

\newcommand{\NN}{\mathbb{N}}
\newcommand{\QQ}{\mathbb{Q}}

\newcommand{\CC}{\mathbb{C}}
\newcommand{\ZZ}{\mathbb{Z}}
\newcommand{\YY}{\mathbb{Y}}

\newcommand{\M}{\overline{\mathcal{M}}}
\newcommand{\I}{\mathcal{I}}
\newcommand{\J}{\mathcal{J}}
\newcommand{\A}{\mathcal{A}}
\newcommand{\F}{\mathcal{F}}

\renewcommand{\H}{\mathcal{H}}

\newcommand{\beq}{\begin{equation}}
\newcommand{\eeq}{\end{equation}}

\newcommand{\nn}{\nonumber}

\newtheorem{dfn}{Definition}[section]
\newtheorem{lem}[dfn]{Lemma}
\newtheorem{prp}[dfn]{Proposition}
\newtheorem{thm}[dfn]{Theorem}

\newtheorem{cor}[dfn]{Corollary}
\newtheorem{emp}[dfn]{Example}

\newenvironment{prf}{\noindent {\it Proof} \ }{\hfill $\Box$}
\newenvironment{prfn}[1]{\noindent {\it Proof of #1.} \ }{\hfill $\Box$}

\DeclareMathOperator{\res}{\mathrm{res}}

\begin{document}

\title{The Hodge-FVH Correspondence}
\author{Si-Qi Liu, \quad Di Yang,\quad Youjin Zhang, \quad Chunhui Zhou}
\date{}
\maketitle

\begin{abstract}
The Hodge-FVH correspondence establishes a relationship between the  
special cubic Hodge integrals and an integrable hierarchy, which is called the 
fractional Volterra hierarchy. In this paper we prove 
this correspondence. As an application of this result, 
we prove a gap condition for certain special cubic Hodge integrals 
and give an algorithm for computing the coefficients 
that appear in the gap condition.
\end{abstract}

\setcounter{tocdepth}{1}
\tableofcontents

\let\thefootnote\relax

\section{Introduction}
The study of the deep relationship between topology of 
the Deligne--Mumford moduli space of stable algebraic curves and integrable hierarchies started from Witten's conjecture~\cite{Witten}.
It states that the formal series 
$Z_{\rm WK}(t;\e)\in \QQ((\e^2))[[t_0,t_1,\dots]]$ defined by  
\begin{align}
& Z_{\rm WK}(t;\e):=\exp \Biggl(\sum_{g\ge0} \e^{2g-2} 
\sum_{k\geq0} \sum_{i_1,\dots,i_k\geq 0} \frac{t_{i_1}\cdots t_{i_k}}{k!} \int_{\M_{g,k}}\psi_1^{i_1}\cdots\psi_k^{i_k} \Biggr)\nn
\end{align} 
is a particular tau-function of the Korteweg--de Vries (KdV) hierarchy. 
Here, $\M_{g,k}$ denotes the Deligne--Mumford moduli space of stable algebraic curves of genus~$g$ with~$k$ distinct 
marked points, $\psi_i$ denotes the first Chern class of the $i_{\rm th}$ cotangent line bundle over $\M_{g,k}$, $i=1,\dots,k$, 
and $t:=(t_i)_{i\geq 0}$ is the infinite vector of indeterminates. 
The first proof of Witten's conjecture was given by Kontsevich in~\cite{Kontsevich}, and several other proofs were given in~\cite{KL, LX,  Mirzak, OP1}.

The Hodge integrals over $\M_{g,k}$ are some rational numbers defined by  
$$
\int_{\M_{g,k}} \psi_1^{i_1}\cdots\psi_k^{i_k}\lambda_{j_1}\cdots\lambda_{j_l},
$$
where $\lambda_j$ is the $j$-th Chern class of the Hodge bundle $\mathbb{E}_{g,k}$ on $\M_{g,k}$, $j=0,\dots,g$.
For the dimension reason these integrals vanish unless 
\beq\label{counting-dim}
i_1+\cdots+i_k+j_1+\cdots +j_l=3g-3+k.
\eeq
They reduce to the $\psi$-class intersection numbers when $j_1=j_2=\dots=0$. 
In this paper we consider 
the generating series of certain cubic Hodge integrals given by
\beq
Z_{\rm cubic}(t;p,q,r;\e)=e^{\sum_{g\ge 0} \e^{2g-2} \H_g(t;p,q,r)},\label{cubicZ}
\eeq
where $\H_g(t;p,q,r)$, $g\geq 0$ are the genus~$g$ cubic Hodge potentials
\beq\label{def-Hg}
\H_g(t; p, q, r)=
\sum_{k\geq0} \sum_{i_1,\dots,i_k\geq 0} \frac{t_{i_1}\cdots t_{i_k}}{k!} \int_{\M_{g,k}}\psi_1^{i_1}\cdots\psi_k^{i_k} 
\mathcal{C}_g(-p)\mathcal{C}_g(-q)\mathcal{C}_g(-r).
\eeq
Here, $\mathcal{C}_g(z):=\sum_{j=0}^g \lambda_j z^j$ denotes the Chern polynomial of $\mathbb{E}_{g,k}$.
It was proved in~\cite{DLYZ-1} that the formal series~$u_{\rm cubic}$ defined by 
\[u_{\rm cubic} := \e^2 \p_{t_0}^2 \log Z_{\rm cubic}(t;p,q,r;\e)\]
satisfies an integrable hierarchy of Hamiltonian PDEs, called the {\it cubic Hodge hierarchy}.

The above cubic Hodge integrals given in \eqref{def-Hg} are called special if $p,q,r$ satisfy the following local Calabi--Yau condition: 
\beq
\frac1p+\frac1q+\frac1r=0.
\eeq
Geometric significance for these special cubic Hodge integrals is manifested in the localization 
technique of computing Gromov--Witten invariants for toric three-folds \cite{Bri,GP, LLZ, OP}, and by the  
Gopakumar--Mari\~no--Vafa conjecture regarding the Chern--Simons/string duality \cite{GV,MV}.

The conjectural Hodge-FVH correspondence~\cite{LZZ}, which is a generalization of Witten's conjecture,
 states that the cubic Hodge hierarchy for the special cubic Hodge integrals is equivalent to the 
{\it fractional Volterra hierarchy} (FVH). The goal of this paper is to prove this conjecture.
To give a precise statement of the Hodge-FVH correspondence, let us introduce some notations. Denote 
\beq
\mathcal{I}_1:=\left\{k p\mid k\in \ZZ_{>0}\right\}, \quad \mathcal{I}_2= \left\{k q \mid k\in \ZZ_{>0}\right\}, 
\quad \mathcal{I} := \mathcal{I}_1 \cup \mathcal{I}_2.
\eeq
The set $\mathcal{I}$ will serve as the set of indices. Also introduce the numbers $c_\mu$, $\mu\in \mathcal{I}$ by 
\begin{align}
& c_\mu := \binom{-\mu/r}{\mu/p}, ~  \mu\in \mathcal{I}_1, \qquad  c_\mu:=\binom{-\mu/r}{\mu/q}, ~  \mu\in \mathcal{I}_2.
\label{constant-0} 
\end{align}
\begin{thm}[Hodge-FVH correspondence]\label{main-1}
Let $\Lambda:=e^{\e \p_x}$.
The formal series 
\beq\label{def-u-cubic}
u(x, T;\e):= \left(\lm^{-\frac{1}{2r}}-\lm^{\frac{1}{2r}}\right)\left(\lm^{\frac1{2q}}-\lm^{-\frac1{2q}}\right)
                  \log Z_{\rm cubic} \biggl(t(x, T); p, q,r; \frac{\sqrt{p+q}}{pq}\e\biggr)
\eeq 
satisfies the FVH. More precisely, denote $L=\lm^{\frac1{p}}+e^u\lm^{-\frac1q}$, then for $\mu\in\I$ we have
\beq\label{FVH-0}
\e\frac{\p L}{\p T_\mu}=
\begin{cases}
\left[\left(L^{-\frac{\mu}{r}}\right)_+,L\right], & \mu \in \mathcal{I}_1 , \\
-\left[\left(L^{-\frac{\mu}{r}}\right)_-,L\right], & \mu \in \mathcal{I}_2 ,
\end{cases}
\eeq
where 
\begin{align}
t_i(x,T):= \frac1{pq}\sum_{\mu\in \mathcal{I}} \mu^{i+1} c_\mu T_\mu -1 + x\delta_{i,0} + \delta_{i,1}. \label{time-trans}
\end{align}
Moreover, the formal series $Z(x,T;\e)\in\CC((\e^2))[[x-1, T]]$ defined by 
\beq\label{defZ}
Z(x,T;\e) := \exp \biggl(\frac{p^2q^2}{(p+q)\e^2} A \biggr)  \,
Z_{\rm cubic} \biggl(t(x,T); p,q,r; \frac{\sqrt{p+q}}{pq}\e\biggr) 
\eeq
is a tau-function of the FVH, where $A$ is an element in $\CC[[x-1,T]]$ defined by
\beq\label{def-a}
A:= 
\frac{1}{2p^2q^2}\sum_{\mu_1,\mu_2\in \I} \frac{\mu_1 \mu_2}{\mu_1 + \mu_2} c_{\mu_1} c_{\mu_2} T_{\mu_1} T_{\mu_2} 
+ \frac{1}{pq} \sum_{\mu \in \I}  \Bigl(x-\frac{\mu}{\mu+1}\Bigr) c_\mu T_\mu + \frac14 - x.
\eeq
\end{thm}

We will give the definition of tau-function of the FVH in Section~\ref{sectiontau}.
Note that for the case $(p,q,r)=(\frac1m,\frac1n,-\frac1{m+n})$ with $m,n\in \NN$, 
the time variables~$T_\mu$ are identified with the time variables~$s_k$ in~\cite{LYZZ}
in the following way:
\[
T_{b_k}=s_k,\quad k\in(\NN-n) \backslash \left(\{0\} \cup ((m+n)\NN-n)\right),
\]
where 
\[
b_{\alpha+(m+n)\ell} := 
\left\{
\begin{array}{cl}
\frac{-\alpha}{n}+\ell,\quad & \alpha=-(n-1),\dots,-1,\\
\frac{\alpha}{m}+\ell,\quad & \alpha=0,\dots,m-1.
\end{array}
\right.
\]

In order to prove Theorem~\ref{main-1}, we will first consider 
the \emph{rational case}, i.e.
\beq\label{pqrmnh}
p=\frac{1}{m},\quad q=\frac{1}{n},\quad r=-\frac{1}{h}, 
\eeq
where $m$ and~$n$ are coprime positive integers, and $h:=m+n$.
The key idea of our proof is to use the uniqueness theorem of 
solutions to the loop equation already shown in~\cite{LYZZ}.
We will show that there exists a particular tau-function $\tau_{\rm top}$ 
of the FVH satisfying certain technical conditions  
required by the uniqueness theorem of~\cite{LYZZ}, 
and then show that $\tau_{\rm top}$ satisfies the Virasoro constraints, 
which leads to the loop equation, and thus to a proof of the theorem.
Theorem~\ref{main-1} in the general case is proved by using a continuation argument.
These proofs are given in Sections~\ref{topologicaltaufunction}--\ref{Proofofmain}. 

As applications of Theorem~\ref{main-1}, we prove 
Propositions \ref{proppri}, \ref{p14} 
and Theorem \ref{gapthm} (see below) 
on properties of certain special 
cubic Hodge integrals. 
Here Proposition \ref{proppri} and Theorem \ref{gapthm} are presented in~\cite{LYZZ} without proof.
For the presentation of Proposition \ref{proppri}, we denote
\beq\label{wHHg}
\widetilde{\mathcal{H}}_g(x, T):=\mathcal{H}_g \bigl(t(x, T); p,q,r \bigr), \quad g\geq 0,
\eeq
where $t(x, T)$ is defined in~\eqref{time-trans}.
\begin{prp}\label{proppri} For the rational case, 
$\widetilde{\mathcal{H}}_g(x, T)$, $g\geq 1$ have the following property:
\begin{align}
& \widetilde{\mathcal{H}}_1(x, T) - \frac{\sigma_1-1}{24} \log x  \label{gap1} \\ 
= & 
\sum_{m\geq 1}\sum_{\lambda_1,\dots,\lambda_m\in \I} \frac{c_{1; \, \lambda_1,\dots,\lambda_m}(\sigma_1,\sigma_3)}{m!} 
x^{\lambda_1+\dots+\lambda_m-m} 
T_{\lambda_1} \cdots T_{\lambda_m},  \nn \\
& \widetilde{\mathcal{H}}_g (x, T) - \frac{R_g(\sigma_1,\sigma_3)}{x^{2g-2}}  \label{gap2} \\ 
=& 
\sum_{m\geq 1}\sum_{\lambda_1,\dots,\lambda_m\in \I}  \!\!\!\!\!
\frac{c_{g; \, \lambda_1,\dots,\lambda_m}(\sigma_1,\sigma_3)}{m!} x^{\lambda_1+\dots+\lambda_m-m-(2g-2)} 
T_{\lambda_1} \cdots T_{\lambda_m}, \quad g\geq 2.\nn
\end{align}
In these formulae, $\sigma_1,\sigma_3$ are related with~$m,n$ by
\beq\label{sigma1sigma3mnh}
\sigma_1=\frac1h-\frac1m-\frac1n,\quad \sigma_3=\frac2{h^3}-\frac2{m^3}-\frac2{n^3},
\eeq
$c_{g; \, \lambda_1,\dots,\lambda_m}(\sigma_1,\sigma_3)$, $g\geq 1$ and $R_g(\sigma_1,\sigma_3)$, $g\geq 2$
are certain polynomials of $\sigma_1,\sigma_3$ satisfying the condition 
\begin{equation}
\deg R_g\leq 3g-3\quad \textrm{with}\  \deg \sigma_1=1, \deg \sigma_3=3,
\end{equation} 
and the degree $(3g-3)$-part of $R_g(\sigma_1,\sigma_3)$ is given by
\beq \label{Faber}
   \frac{(-1)^g}{2(2g-2)!}  \frac{|B_{2g}||B_{2g-2}|}{2g (2g-2)} \left(\frac{\sigma_1^3}3-\frac{\sigma_3}{6}\right)^{g-1},
\eeq
where $B_j$ denotes the $j_{\rm th}$ Bernoulli number.
\end{prp}
The proof of this proposition is given in Section~\ref{GAP}.

The coefficients $R_g(\sigma_1,\sigma_3 )$ appearing in the formula~\eqref{gap2} are hard to 
compute. In \cite{LYZZ} we gave the expressions of~$R_2$ and~$R_3$. 
Based on the Hodge-FVH correspondence, in particular on the 
introduction of the previously mentioned 
tau-function~$\tau_{\rm top}$ of the FVH,
we will give a new algorithm for computing~$R_g$.
Indeed, the $\tau_{\rm top}$ corresponds to a solution $u_{\rm top}(x,T;\e)$ to the FVH, 
whose initial value $V(x;\e)=u_{\rm top}(x, T=0;\e)\in \QQ[[x-1;\e]]$ satisfies the following difference equation:
\beq
\sum_{0\leq\alpha_1\leq\cdots\leq\alpha_m\leq n} 
e^{\sum_{j=1}^{m} V\left(x+\alpha_j m\e- \left(j-\frac12\right) n\e;\e\right)} =
\binom{m+n}m x . \label{eq-initial} 
\eeq
We will show in Section~\ref{topologicaltaufunction} that this 
equation has a unique solution in $\CC[[x-1;\e]]$. We will also show that 
this unique solution has the form
\beq\label{initial-form}
V(x;\e)=\frac1m\log x+\sum_{g\geq1}\e^{2g} \frac{P_g(m,n)}{x^{2g}},
\eeq
where $P_g(m,n)\in\QQ[m, n, m^{-1}]$. 
Define a sequence of rational numbers $C_k(m,n)$, $k\geq 0$ by 
the generating function
\beq\label{def-Cg}
\sum_{k\geq0}C_k(m,n) z^{2k}
=\frac{nh z^2}{4 \sinh\left(\frac {n z}2\right) \sinh\left(\frac{h z}2\right)}
=1-\frac{n^2+h^2}{24}z^2+\frac{7n^4+10 n^2 h^2+7 h^4}{5760}z^4+\cdots.
\eeq 
These rational numbers have the following explicit expression:
\beq\label{expr-Cg}
C_k(m,n)=
	\sum_{i=0}^k\frac{\left(1-2^{1-2i}\right)\left(1-2^{1-2k+2i}\right)}{(2i)!(2k-2i)!}B_{2i}B_{2k-2i} n^{2i} h^{2k-2i}.
\eeq
\begin{prp}\label{p14}
The polynomials $R_g(\sigma_1,\sigma_3)$ in Proposition~\ref{proppri}
can be represented in terms of~$P_g(m,n)$  and $C_k(m, n)$ as follows:
\begin{equation}
R_g
=\frac{(2g-3)!}{(mnh)^{g}}\left(m\sum_{k=1}^{g}\frac{C_{g-k}(m,n) P_k(m,n)}{(2k-1)!}-C_g(m,n)\right),\quad g\geq2.\label{P-2}
\end{equation}
\end{prp}

By using the Hodge-FVH correspondence we also have the following 
theorem on the gap condition satisfied by certain special cubic Hodge
integrals.
\begin{thm}\label{gapthm}
For the rational case satisfying the condition that one of $p,q$ is equal to~1, 
the free energies 
$\widetilde{\mathcal{H}}_g(x,T)$, $g\geq 1$
 satisfy the following gap condition:
\begin{align}
&\widetilde{\mathcal{H}}_1(x, T)|_{T_{\rm II}=0} - \frac{\sigma_1-1}{24} \log x \in \CC[[x,T_{\rm I}]],  
\label{stronggap1}\\
&\widetilde{\mathcal{H}}_g(x,T)|_{T_{\rm II}=0} - \frac{R_g(\sigma_1,\sigma_3)}{x^{2g-2}} \in \CC[[x,T_{\rm I}]], 
\quad g\geq 2. \label{stronggap2}
\end{align} 
Here $T_{\rm I}= (T_1,T_{2},T_{3},\dots)$ and $T_{\rm II}= (T_\mu)_{\mu\in \mathcal{I}\setminus\mathbb{N}} $.
\end{thm}
The proof of the above theorem is given in Section~\ref{GAP}.

\paragraph{Organization of the paper.} In Section~\ref{sectiontau} 
we give two definitions of tau-functions of the FVH,
and prove the equivalence of these definitions.
In Sections~\ref{topologicaltaufunction}--\ref{Proofofmain} we prove Theorem~\ref{main-1}.
In Section~\ref{GAP}, as applications of Theorem~\ref{main-1}, 
we give a proof of Proposition~\ref{proppri} as well as 
a proof of the gap phenomenon for certain special cubic Hodge integrals described by Theorem \ref{gapthm}.

\paragraph{Acknowledgements.} 
We are grateful to Boris Dubrovin for his support of the work and very helpful discussions. 
We would like to thank Shuai Guo, Yongbin Ruan and Don Zagier for their interests 
and helpful comments.
This work is partially supported by NSFC No.\,11771238, No.\,11725104, No.\,11671371.
Part of the work of C.~Zhou was done during his PhD studies at Tsinghua University; he thanks Tsinghua University for 
excellent working conditions and financial supports.

\section{Tau-functions of the FVH}\label{sectiontau}
In this section, we first recall the definition of the tau-function of an arbitrary 
solution $u(x, T;\e)$ to the FVH that is given in~\cite{LZZ}. 
Then we give an alternative definition, following the ones for the KP and 2D-Toda hierarchies \cite{DKJM, Sato, UT}, 
of the tau-function in terms of wave functions of the integrable hierarchy, and show the equivalence of the two definitions. 
We will use the first definition of the tau-function in Section~\ref{topologicaltaufunction} to construct 
the topological tau-function $\tau_{\rm{top}}$ of the FVH, and to prove that it has the form of the genus expansion required for the derivation of the loop equation \cite{LYZZ} from the Virasoro constraints. The second definition will be used in Section 4 to prove that 
the topological tau-function $\tau_{\rm{top}}$ satisfies the Virasoro constraints.

Let us introduce some notations. We denote by $\A_u$ the ring $\A_{u,0}[u_x,u_{xx},\dots]$ of differential polynomials of~$u$, 
where $\A_{u,0}$ denotes the ring of smooth functions of~$u$. 
Note that the shift operator~$\Lambda=e^{\e \p_x}$ acts on~$\A_u[[\e]]$.
Introduce a gradation ${\rm deg}$ on $\A_u[[\e]]$, defined via the following degree assignments:
\beq\label{grading}
{\rm deg} \, \p_x^k u = k, \quad {\rm deg} \, \e =-1, 
\eeq
where $k\geq 1$. 
For convenience, we denote
\[
\lm_1=\lm^{\frac1q},\quad \lm_2=\lm^{\frac1p},\quad \lm_3=\lm^{\frac1p+\frac1q}.
\] 
Then the operator~$L$ introduced in Theorem~\ref{main-1}
has the form
\beq
L=\lm_2+e^u \lm_1^{-1}.\label{zh-1}
\eeq
We also extend the gradation ${\rm deg}$ to the difference operators of the form 
$\sum_{j\in \ZZ} g_j \Lambda_3^j \Lambda^a$
with $g_j\in \A_u[[\e]]$, $a\in \CC$ by assigning ${\rm deg} \, \Lambda=0$. 

\subsection{The definition of the  tau-function}\label{zh-3}
The following two-point correlation functions in~$\A_u[[\e]]$ were introduced in~\cite{LZZ}:
\begin{align} \label{defomega}
\Omega_{\lambda,\mu}= \left\{\begin{array} {cl}
\Lambda_2^\frac12 \sum_{j=1}^{\lambda/p}\frac{\lm_3^j-1}{\lm_3-1}
\left(\res_{\lm_3} \left(\lm_3^{-j} L^{-\frac{\lambda}r}\right) \res_{\lm_3}\left(L^{-\frac{\mu}r} \lm_3^{j}\right)\right),
& \lambda\in \mathcal{I}_1,\\
\Lambda_2^\frac12 \sum_{j=1}^{\lambda/q}\frac{\lm_3^j-1}{\lm_3-1}
\left(\res_{\lm_3} \left( L^{-\frac{\lambda}r} \lm_3^j \right) \res_{\lm_3}\left(\lm_3^{-j}L^{-\frac{\mu}r}\right)\right),
& \lambda\in \mathcal{I}_2, 
\end{array}\right. 
\end{align}
where $\mu \in \mathcal{I}$, and $\res_{\Lambda_3}$ means to take ``residue" in the following sense:
\beq
\res_{\lm_3}\sum_{i\in\ZZ}a_i\lm_3^i:=a_0.
\eeq
In the above formula~\eqref{defomega}, $L^{-\frac{\mu}r}\in \A_u[[\e]]((\Lambda_3^{-1}))$ 
when $\mu\in \I_1$, and  
$L^{-\frac{\mu}r}\in \A_u[[\e]]((\Lambda_3))$ when $\mu\in \I_2$. 
See in~\cite{LZZ} for the precise 
definition of these fractional powers of~$L$.

\begin{prp}[\cite{LZZ}]\label{taustructure}
Let $u=u(x,T;\e)$ be an arbitrary solution in $\CC\left[\left[x-1,T;\e\right]\right]$ to the FVH~\eqref{FVH-0}. 
Then the following formulae hold true for any $\lambda,\mu,\nu\in \mathcal{I}$:
\begin{align}
& \Omega_{\lambda,\mu} = \Omega_{\mu,\lambda}, \quad \frac{\p \Omega_{\lambda,\mu}}{\p T_{\nu}} = \frac{\p \Omega_{\mu,\nu}}{\p T_{\lambda}}, 
\label{ooproperty}\\
& (\lm_3-1) \lm_2^{-\frac12}\Omega_{\lambda,\mu}=\e\frac{\p}{\p T_{\lambda}}\res_{\lm_3}L^{-\frac{\mu}r}.
\label{orproperty}
\end{align}
\end{prp}

It follows from Proposition~\ref{taustructure} that 
for an arbitrary solution~$u=u(x,T;\e)$ in $\CC\left[\left[x-1,T;\e\right]\right]$ 
to the FVH,
there exists $\tau=\tau(x,T;\e)\in\CC((\e))[[x-1, T]]$ satisfying the relations
\begin{align}
&u=\left(\lm_3^\frac12-\lm_3^{-\frac12}\right)\left(\lm_1^{\frac12}-\lm_1^{-\frac12}\right)\log \tau,\label{taudef1}\\
&\e\left(\lm_3-1\right)\lm_2^{-\frac12}\frac{\p \log \tau}{\p T_\lambda}=\res_{\lm_3} L^{ - \frac{\lambda}{r}},\quad \lambda \in \mathcal{I},\label{taudef2}\\
&\e^2 \frac{\p^2 \log \tau}{\p T_{\lambda} \p T_{\mu}}=\Omega_{\lambda,\mu},\quad \lambda,\mu\in \mathcal{I}. \label{taudef3}
\end{align}
We call~$\tau$ the tau-function of the solution~$u$ to the FVH.
Note that the tau-function~$\tau$ is 
uniquely determined by~\eqref{taudef1}--\eqref{taudef3} up to multiplying by
a factor of the form
\begin{equation}
\exp\biggl\{a x+b+\sum_{\lambda\in\mathcal{I}} d_\lambda T_\lambda\biggr\}, \label{explinfac}
\end{equation}
where $a, b, d_\lambda\in \CC((\e))$ are arbitrary constant Laurent series in~$\e$.

\begin{lem}\label{leading-omega}
The following relations hold true for $\lambda,\mu\in\I$:
\begin{align}
& \res_{\Lambda_3} L^{-\lambda/r}-c_\lambda e^{\lambda u/ p}\in \e\A_u[[\e]],\label{2-10}\\
& \Omega_{\lambda,\mu}- \frac1{p+q} 
\frac{\lambda\mu}{\lambda+\mu} c_\lambda c_\mu e^{(\lambda+\mu)u/p}\in \e\A_u[[\e]],\label{2-11}\\
& \deg \res_{\Lambda_3} L^{-\lambda/r}=0,\quad
{\rm deg} \, \Omega_{\lambda,\mu} = 0.\label{2-12}
\end{align}
Here the constants $c_\lambda, c_\mu$ are defined in \eqref{constant-0}.
\end{lem}
\begin{prf}
From the definition of the fractional powers of~$L$ given in \cite{LZZ} we have, for any $k\in \ZZ$, 
\begin{align}
&\res_{\lm_3} L^{-\lambda/r}\lm_3^{-k}-\binom{-\lambda/r}{\lambda/p-k}e^{\left(\lambda/p-k\right)u}
\in \e\A_u[[\e]],\quad \lambda\in \I_1, \label{leading-coef-L1} \\
&\res_{\lm_3} L^{-\lambda/r}\lm_3^{-k}-\binom{-\lambda/r}{\lambda/q+k}e^{\left(\lambda/p-k\right)u}
\in \e\A_u[[\e]],\quad \lambda\in \I_2,  \label{leading-coef-L2} \\
&
{\rm deg}\,\res_{\lm_3} L^{-\lambda/r}\lm_3^{-k}=0,\quad \lambda\in\I .
\end{align}
Then by using the definition~\eqref{defomega} of~$\Omega_{\lambda,\mu}$ we find, for any $\lambda,\mu\in\I_1$, that 
${\rm deg}\,\Omega_{\lambda,\mu}=0$  and 
\begin{align*}
&\Omega_{\lambda,\mu}
-\frac1{p+q} \frac{\lambda\mu}{\lambda+\mu} c_\lambda c_\mu e^{(\lambda+\mu)u/p} = \Omega_{\lambda,\mu}
-\sum_{j=1}^{\lambda/p}j\binom{-\lambda/r}{\lambda/p-j}\binom{-\mu/r}{\mu/p+j}e^{(\lambda+\mu)u/p}
\in\e\A_u[[\e]].
\end{align*}
In the above equality, the following elementary identity (cf.~\cite{LZZ}) is used:
\[
\sum_{j=1}^{\lambda/p}j\binom{-\lambda/r}{\lambda/p-j}\binom{-\mu/r}{\mu/p+j}
=\frac1{p+q} \frac{\lambda\mu}{\lambda+\mu} c_\lambda c_\mu,\quad \lambda,\mu\in\I_1. 
\]
This proves \eqref{2-10}--\eqref{2-12} for the case $\lambda,\mu\in\I_1$. 
The proof for the rest cases is similar.  
\end{prf}

\subsection{An alternative definition of the  tau-function}\label{zh-2}
The tau-function of a solution to the FVH can also be defined, following the approach of \cite{DKJM, Sato, UT}, via the introduction of wave functions for the FVH.
To this end, we first introduce the dressing operators.

\begin{lem}\label{dressing}
For an arbitrary solution $u=u(x, T;\e)$ in $\CC[[x-1,T;\e]]$ to the FVH~\eqref{FVH-0}, there exist difference operators $\Phi_1,\Phi_2$,
called the dressing operators, of the form
\beq
\Phi_1(x, T;\e)=\sum_{i\geq 0} a_{1,i}(x, T;\e)\lm_3^{-i},\qquad
\Phi_2(x, T;\e)=\sum_{i\geq 0} a_{2,i}(x, T;\e)\lm_3^i \label{Phiform}
\eeq
with $a_{1,i},a_{2,i} \in \CC((\e))[[x-1, T]]$, $i\geq 0$ and $a_{1,0}(x, T;\e)\equiv1$, such that 
\begin{align}
& L=\Phi_1 \lm_2 \Phi_1^{-1}=\Phi_2 \lm_1^{-1} \Phi_2^{-1},  \label{Ldecom}\\
& \e\frac{\p \Phi_1}{\p T_\mu}=-\left(\Phi_1 \lm_3^{\mu/p} \Phi_1^{-1}\right)_-\Phi_1, \quad 
\e\frac{\p \Phi_2}{\p T_\mu}= \left(\Phi_1 \lm_3^{\mu/p} \Phi_1^{-1}\right)_+\Phi_2,\quad \mu \in\I_1,  \label{dphisk1}\\
& \e\frac{\p \Phi_1}{\p T_\mu}=-\left(\Phi_2 \lm_3^{-\mu/q} \Phi_2^{-1}\right)_-\Phi_1, \quad 
 \e\frac{\p \Phi_2}{\p T_\mu}= \left(\Phi_2 \lm_3^{-\mu/q} \Phi_2^{-1}\right)_+\Phi_2,\quad \mu \in\I_2.  \label{dphisk2}
\end{align}
These two operators $\Phi_1$ and $\Phi_2$ are uniquely determined by~$u$ up to right-multiplications by operators of the form
\[
\Phi_1\mapsto \Phi_1\left(1+c_{1,1}\lm_3^{-1}+c_{1,2}\lm_3^{-2}+\cdots\right),\quad
\Phi_2\mapsto \Phi_2\left(c_{2,0}+c_{2,1}\lm_3+c_{2,2}\lm_3^2+\cdots\right),
\]
where $c_{i,j}$ are arbitrary Laurent series of~$\e$ with constant coefficients. 
\end{lem}
\begin{prf}
The equation $L=\Phi_1\lm_2 \Phi_1^{-1}$ is equivalent to the following recursion relations:
\begin{align*}
&\left(\lm_2-1\right)a_{1,1}=-e^u,\\
&\left(\lm_2-1\right)a_{1,i}=-e^u\lm_1^{-1}a_{1,i-1},\quad i\geq2.
\end{align*}
On another hand, by using the FVH~\eqref{FVH-0} one can check that 
equations \eqref{Ldecom}--\eqref{dphisk2} for~$\Phi_1$ are compatible, 
from which we know the existence of~$\Phi_1$.
If we have another operator $\Phi_1'$ which 
satisfies the same equations as for $\Phi_1$, then $\Phi_1^{-1}\Phi_1'$ satisfies the following two equations:
$$
\big[\Phi_1^{-1}\Phi'_1,\,\lm_2\big]=0,\quad
\e\p_{T_\mu}\big(\Phi_1^{-1}\Phi'_1\big)=0,\quad \forall\,\mu\in \I.
$$
Hence $\Phi_1^{-1}\Phi'_1$ must have the form
\[
\Phi_1^{-1}\Phi'_1=1+c_{1,1}\lm_3^{-1}+c_{1,2}\lm_3^{-2}+\cdots. 
\]
The assertion of the lemma for~$\Phi_2$ can be proved in a similar way.
\end{prf}

Given a solution $u=u(x, T;\e)\in\CC[[x-1, T;\e]]$ 
to the FVH~\eqref{FVH-0}, we have the associated wave functions
$\psi_1$, $\psi_2$ defined by
\beq\label{psiphi}
\psi_1:=\Phi_1 e^{\vartheta_1}, \quad  
\psi_2:=\Phi_2 e^{\vartheta_2},
\eeq
where $\Phi_1$, $\Phi_2$ are the dressing operators,
\begin{align}
&\vartheta_1=\vartheta_1(x, T; \e; z)
:=-r\frac x \e \log z + \sum_{\mu \in \J_1} \frac{T_\mu}\e z^{\frac{\mu}p}  + \frac12 \sum_{\mu \in \J_3} \frac{T_\mu}\e z^{\frac{\mu}p},\label{zh-19}\\
&\vartheta_2=\vartheta_2(x, T; \e; z)
:=r\frac x \e \log z-\sum_{\mu \in \J_2} \frac{T_\mu}\e z^{\frac{\mu}q}  
- \frac12 \sum_{\mu \in \J_3} \frac{T_\mu}\e z^{\frac{\mu}q},\label{zh-20}
\end{align}
and the index sets $\J_1$, $\J_2$, $\J_3$ are defined by
\beq\label{zh-18}
\J_1:=\I_1 \backslash \I_2, \quad \J_2:=\I_2 \backslash \I_1, \quad \J_3 := \I_1 \cap \I_2. 
\eeq
It is easy to verify that $\psi_1,\psi_2$ satisfy the following equations:
\beq\label{defwavefunctions}
L\psi_i=\lambda_i \psi_i,\quad \e\frac{\p \psi_i}{\p T_\mu}=P_\mu \psi_i,\quad \mu\in\mathcal{I}, ~ i=1,2.
\eeq
Here $\lambda_1=z^{-r/p}$, $\lambda_2=z^{-r/q}$, and
\beq
P_{\mu} := \left\{
\begin{array}{cl}
\left(L^{-\frac\mu{r}}\right)_+, & \mu \in \J_1,            \\
-\left(L^{-\frac\mu{r}}\right)_-, & \mu \in \J_2,             \\
 \frac{1}{2}\left(L^{-\frac\mu{r}}\right)_+ -\frac{1}{2}\left(L^{-\frac\mu{r}}\right)_-, & \mu \in \J_3.
\end{array}\right.
\eeq

We also define the dual wave functions $\psi_1^*$ and $\psi_2^*$ by 
\beq
\psi^*_1:= \left(\Phi_1^{-1}\right)^*e^{-\vartheta_1}, \quad \psi^*_2 := \left(\Phi_2^{-1}\right)^*e^{-\vartheta_2}.
\eeq
Here we recall that $\lm^*:=\lm^{-1}$, 
and for two difference operators $A$ and $B$, 
\beq
 \quad \left(A B\right)^*:=B^* A^*.
\eeq

\begin{thm}\label{bilineareq}
A formal series~$u\in\CC[[x-1, T;\e]]$ satisfies the FVH if and only if the associated wave functions 
and their duals satisfy the following bilinear equations for all $\ell\in \ZZ$:
\begin{align} \label{bilinearid12}
\res_z \psi_1(x, T; \e; z) \, \psi_1^*\Bigl(x - \frac \ell r\e, T'; \e; z\Bigr)\frac{dz}{z}  =
\res_z \psi_2(x, T;\e;z) \, \psi_2^*\Bigl(x-\frac \ell r \e, T';\e;z\Bigr)\frac{dz}{z}.
\end{align}
\end{thm}

\begin{lem}\label{d1equaltod2}
Two operator-valued formal series of the form
\[
D_1(x, T)\in \CC\left(\left(\lm_3^{-1},\e\right)\right)[[x-1, T]],\quad 
D_2(x, T)\in \CC\left(\left(\lm_3,\e\right)\right)[[x-1, T]]
\] 
are equal if and only if the following identity holds true for all $\ell\in \ZZ$:
\begin{align}
& \res_z D_1(x, T) \left(\psi_1(x, T;z)\right) \lm_3^\ell \left(\psi_1^*(x, T';z)\right)\frac{dz}{z} \\
& \qquad = \res_z D_2(x, T)\left( \psi_2(x, T;z)\right) \lm_3^\ell\left(\psi_2^*(x, T';z)\right)\frac{dz}{z}. \nn
\end{align}
\end{lem}
The proofs of the above Theorem~\ref{bilineareq} and Lemma~\ref{d1equaltod2} are standard (cf.~\cite{ASM}).

Let us now introduce the notion of tau-function via the wave functions. The following notations will be used.
Denote by 
$\phi_i$ and $\phi_i^*$, $i=1,2$ 
the following formal series:
\begin{align}
& \phi_1(x, T; \e; z):=\Phi_1(x, T;\e)|_{\Lambda_3\to z}, \quad 
\phi_2(x, T; \e; z):=\Phi_2(x, T;\e)|_{\Lambda_3\to z^{-1}}, \label{phiPhi1} \\
& \phi_1^*(x, T; \e; z):=\Phi_1^*(x, T;\e)|_{\Lambda_3\to z^{-1}}, \quad 
\phi_2^*(x, T; \e; z):=\Phi_2^*(x, T;\e)|_{\Lambda_3\to z}.  \label{phiPhi2}
\end{align}
Clearly, we have
\[\phi_1= \sum_{i\geq 0} a_{1,i}(x, T;\e) \, z^{-i}, \quad 
\phi_2=\sum_{i\geq 0} a_{2,i}(x, T;\e) \, z^{-i}.\]
For an arbitrary function~$f( T)$, denote
\begin{align*}
& f \bigl( T - \left[z^{-1}\right]_1\bigr)
=\exp\biggl(-\sum_{\mu \in \I_1}\frac{z^{-\mu/p}}{\mu/p}\e\frac{\p}{\p T_\mu} \biggr) f(T),\\
& f \bigl( T- \left[z^{-1}\right]_2\bigr)
=\exp\biggl(-\sum_{\mu \in \I_2}\frac{z^{-\mu/q}}{\mu/q}\e\frac{\p}{\p T_\mu} \biggr) f(T).
\end{align*}

\begin{thm}\label{wave-tau}
For an arbitrary solution $u(x, T;\e)$ in $\CC[[x-1, T;\e]]$ to the FVH, 
there exists a choice of the dressing operators $\Phi_i$, $i=1,2$ such that the associated formal series 
$\phi_i,\phi_i^*$, $i=1,2$ can be 
represented in terms of a certain formal power series $\tau\in\CC((\e))[[x-1, T]]$ as follows:
\begin{align}
\phi_1(x, T;\e;z)=\frac{\tau(x, T-[z^{-1}]_1;\e)}{\tau(x, T;\e)},\quad
\phi_2(x, T;\e;z)=\frac{\tau(x-\e/r, T-[z^{-1}]_2;\e)}{\tau(x, T;\e)},\label{tau2}\\
\phi^*_1(x, T;\e;z)=\frac{\tau(x-\e/r, T+[z^{-1}]_1;\e)}{\tau(x-\e/r, T;\e)},\quad
\phi^*_2(x, T;\e;z)=\frac{\tau(x, T+[z^{-1}]_2;\e)}{\tau(x-\e/r, T;\e)}.\label{tau2s}
\end{align}
We call $(\Phi_1,\Phi_2,\tau)$ a dressing triple for the FVH. We also call
 the formal power series~$\tau$ the Sato type tau-function of the solution~$u$. 
\end{thm}
\begin{prf} We prove the theorem by employing the method used in~\cite{UT}.
Firstly, by using the bilinear equations~\eqref{bilinearid12} one can prove that 
\begin{align}
&\phi_1(x, T;\e;\zeta) \, 
\phi^*_1\left(x+\frac{\e}r, T-[\zeta^{-1}]_1-\left[\xi^{-1}\right]_1;\e;\zeta\right) \label{ww11} \\
&\qquad = \phi_1(x, T;\e;\xi) \,
\phi_1^*\left(x+\frac{\e}r, T-[\zeta^{-1}]_1-\left[\xi^{-1}\right]_1; \e; \xi \right), \nn\\
&\phi_1(x, T;\e;\zeta) \, 
\phi_1^*\left(x, T-[\zeta^{-1}]_1-\left[\xi^{-1}\right]_2;\e;\zeta\right) \label{ww12} \\
&\qquad = \phi_2(x, T;\e;\xi) \,
\phi_2^*\left(x, T-[\zeta^{-1}]_1-\left[\xi^{-1}\right]_2;\e;\xi\right), \nn \\
&\phi_2(x, T;\e;\zeta) \, 
\phi_2^*\left(x-\frac{\e}r, T-[\zeta^{-1}]_2-\left[\xi^{-1}\right]_2;\e;\zeta\right)  \label{ww22} \\
&\qquad = \phi_2(x, T;\e;\xi) \, 
\phi_2^*\left(x-\frac{\e}r, T-[\zeta^{-1}]_2-\left[\xi^{-1}\right]_2;\e;\xi\right). \nn
\end{align}
Comparing the coefficients of~$\zeta^0$ 
of both sides of~\eqref{ww11} and of~\eqref{ww12} respectively, 
we obtain
\beq\label{dual-tau}
\phi_1(x, T;\e;\xi) \, \phi_1^*\Bigl(x+\frac{\e}r, T-\left[\xi^{-1}\right]_1;\e;\xi\Bigr)=1,
\quad
\phi_2(x, T;\e;\xi) \, \phi_2^*\Bigl(x, T-\left[\xi^{-1}\right]_2;\e;\xi\Bigr)=1.
\eeq
By using these two relations we can eliminate~$\phi_1^*$ 
and~$\phi_2^*$ in~\eqref{ww11}--\eqref{ww22} and obtain
\begin{align}
&\phi_1\left(x, T- \left[\xi^{-1}\right]_1;\e;\zeta\right) \phi_1\left(x, T;\e;\xi\right)
= \phi_1\left(x, T- \left[\zeta^{-1}\right]_1;\e;\xi\right) \phi_1\left(x, T;\e;\zeta\right), \label{ww1} \\
&\phi_1\left(x-\frac{\e}r, T- \left[\xi^{-1}\right]_2;\e;\zeta\right) \phi_2\left(x, T;\e;\xi\right)
= \phi_2\left(x, T-[\zeta^{-1}]_1;\e;\xi\right) \phi_1\left(x, T;\e;\zeta\right),\label{ww2}\\
&\phi_2\left(x-\frac{\e}r, T-\left[\xi^{-1}\right]_2;\e;\zeta\right)
 \phi_2\left(x, T;\e;\xi\right)
=\phi_2\left(x-\frac{\e}r, T-[\zeta^{-1}]_2;\e;\xi\right)
 \phi_2\left(x, T;\e;\zeta\right).\label{ww3}
\end{align}
Equations~\eqref{ww1} and~\eqref{ww3} imply the existence of two formal series 
\[\tau_1, \tau_2\in\CC((\e))[[x-1, T]]\] 
satisfying the equations
\[
\phi_1(x, T;\e;z)
=\frac{\tau_1(x, T-[z^{-1}]_1;\e)}{\tau_1(x, T;\e)},
\quad
\phi_2(x, T;\e;z)
=\frac{\tau_2(x-\e/r, T-[z^{-1}]_2;\e)}{\tau_2(x, T;\e)}.
\]
Define $f=\log\tau_2-\log\tau_1$, then we obtain from equation~\eqref{ww2} that 
\begin{align*}
&f\left(x-\frac\e{r},  T-\left[\xi^{-1}\right]_1-[\zeta^{-1}]_2; \e\right)+f\left(x,  T; \e\right)\\
=&
f\left(x-\frac\e{r},  T-[\zeta^{-1}]_2; \e\right)+f\left(x,  T-\left[\xi^{-1}\right]_1; \e\right),
\end{align*}
whose general solution in $\CC((\e))[[x-1, T]]$ has the form
\[f(x,  T; \e)=f_1\left(x,  T''; \e\right)+f_2 \left( T'; \e\right)+f_0( T'''; \e)+b(\e),\]
where
\begin{align*}
 T' =\left(T_\mu\mid \mu \in \J_1\right), \quad
 T''=\left(T_\mu\mid \mu \in \J_2\right), \quad
 T'''=\left(T_\mu\mid \mu \in \J_3\right),
\end{align*}
and $f_0\in\CC((\e))[[ T''' ]]$ is linear in~$T'''$,
\[f_1\left(x,  T''; \e\right)\in\CC((\e))[[x-1, T'']], 
\quad f_2\left( T'; \e\right)\in\CC((\e))[[ T']],
\quad b(\e)\in\CC((\e)).\]
Let $\tau=\tau_1 e^{f_1}$, then we have
\begin{align*}
&\log\phi_1\left(x, T;\e;z\right)
=\log\tau\left(x, T-[z^{-1}]_1;\e\right)-\log\tau\left(x, T;\e\right),\\
&\log\phi_2\left(x, T;\e;z\right)
=\log\tau\left(x-\frac{\e}r, T-[z^{-1}]_2;\e\right)-\log\tau\left(x, T;\e\right)-\log C\left(\e;z\right)
\end{align*}
for some $C(\e;z)\in\CC[[\e;z^{-1/q}]]$.

Finally, by replacing $\Phi_2$ with $\Phi_2 \circ C\bigl(\e;\lm_3^{-1}\bigr)$, we find that 
$\Phi_1$, the new~$\Phi_2$, and~$\tau(x, T;\e)$ satisfy the equations given in~\eqref{tau2}.
The equations given in~\eqref{tau2s} then follow from~\eqref{dual-tau}. 
The theorem is proved.
\end{prf}

We note that for a fixed solution $u\in\CC[[x-1,T;\e]$ the Sato type 
tau-function $\tau$ of the solution is uniquely 
determined up to a factor of the form~\eqref{explinfac}.

\subsection{The equivalence of the two definitions of the tau-function}
For an arbitrary solution~$u$ in $\CC[[x-1, T;\e]]$ to the FVH, 
let $\tau_{\mathrm{s}}$
be the tau-function defined in Section~\ref{zh-2}.
We prove in this subsection the equivalence of $\tau_{\mathrm{s}}$ with the tau-function defined in Section~\ref{zh-3}. 
In other words, we are to prove that $\tau_{\mathrm{s}}$ satisfies the relations \eqref{taudef1}--\eqref{taudef3}.
Let us start with the proof of the following lemma.
\begin{lem}\label{dxdt-tau}
The following formula holds true:
\beq
\e(\lm_3-1)\frac{\p\log\tau_{\mathrm{s}}}{\p T_\mu}
=\res_{\lm_3}L^{-\mu/r},\quad \mu\in\I.
\eeq
\end{lem}
\begin{prf}
Let $\Phi_1, \Phi_2$ be the dressing operators for the FVH such that $\Phi_1,\Phi_2,\tau_{\mathrm{s}}$ form a 
dressing triple. 
Recall that $\Phi_2=\sum_{i\geq0}a_{2,i}\lm_3^i$. 
By using the second formula of~\eqref{tau2} we obtain
\beq\label{tau-a}
(\lm_3-1)\log\tau_{\mathrm{s}}(x, T;\e)=\log a_{2,0}(x, T;\e).
\eeq
Also, it follows from the equations in \eqref{dphisk1}, \eqref{dphisk2} that
\beq\label{theaboveformula}
\e\frac{\p a_{2,0}}{\p T_\mu}=\left(\res_{\lm_3}L^{-\mu/r}\right)a_{2,0}.
\eeq
The lemma is then proved by substituting~\eqref{tau-a} into~\eqref{theaboveformula}.
\end{prf}

\begin{lem}\label{2pt-1}
$\forall\,\lambda,\mu\in \I$, the following relations hold true:
\begin{align}
& \e^2\frac{\p^2 \log\tau_{\mathrm{s}}}{\p T_\lambda \p T_\mu} \in \A_u[[\e]], \label{diffpoly1}\\
& {\rm deg} \, \e^2\frac{\p^2 \log\tau_{\mathrm{s}}}{\p T_\lambda \p T_\mu} = 0. \label{diffpoly2}
\end{align}
\end{lem}
\begin{prf}
Let  
$\psi_1,\psi_2$ be the wave functions associated to $\Phi_1,\Phi_2$ respectively.
It follows from \eqref{psiphi}, \eqref{phiPhi1} that 
\beq\label{dtphi-0}
\e \frac{\p_{T_\mu}\phi_1}{\phi_1} 
= \e \frac{\p_{T_\mu}\psi_1}{\psi_1}-\e\p_{T_\mu}\vartheta_1
=\frac{P_\mu \psi_1}{\psi_1}-\e\p_{T_\mu}\vartheta_1
=- \frac1{\psi_1} \left(L^{-\mu/r}\right)_-\psi_1, \quad \mu\in \I.
\eeq
Observe that the operator $\left(L^{-\mu/r}\right)_-$ can be represented in the form
\beq\label{expansion-1}
\left(L^{-\mu/r}\right)_-= - \sum_{k\geq1}f_{\mu,k}(u,u_x,\dots)L^{kp/r},
\eeq
where 
$f_{\mu,k}=f_{\mu,k}(u,u_x,\dots;\e)\in \A_u[[\e]]$ 
are given by the following recursion relations:
\beq\label{f-recursion}
f_{\mu,k}=-\res_{\lm_3}\left(L^{-\mu/r}\lm_3^k\right)
-\sum_{j=1}^{k-1}f_{\mu,j}\res_{\lm_3}\left(L^{jp/r}\lm_3^k\right),\quad k\geq1.
\eeq
Then from Lemma~\ref{leading-omega} it follows that
${\rm deg}\,f_{\mu,k}=0$.
By using the relation $L^{-p/r}\psi_1=z\psi_1$ we find that the formula~\eqref{dtphi-0} can be written as
\beq\label{dtphi-1}
\e \frac{\p_{T_\mu}\phi_1}{\phi_1}
=\sum_{k\geq1}f_{\mu,k}z^{-k}\in\A_u[[\e]][[z^{-1}]],\quad \mu \in \I.
\eeq
From Theorem~\ref{wave-tau} we know that 
\begin{align}
\e\frac{\p_{T_\mu}\phi_1}{\phi_1}
&=-z^{-1} \e^2 \p_{T_\mu}\p_{T_p}\log\tau_{\mathrm{s}} 
- \frac{z^{-2}}2 \left(\e^2\p_{T_\mu}\p_{T_{2p}}\log\tau_{\mathrm{s}}-\e^3\p_{T_\mu}\p^2_{T_p}\log\tau_{\mathrm{s}}\right) +\cdots.\label{dtphi-2}
\end{align}
Then the relations~\eqref{diffpoly1}--\eqref{diffpoly2} for $\lambda\in \I_1$, $\mu\in \I$ follow from~\eqref{dtphi-1} and~\eqref{dtphi-2}. 
The relations \eqref{diffpoly1}--\eqref{diffpoly2} for $\lambda\in\I_2$, $\mu\in \I$ can be proved in a similar way. 
The lemma is proved.
\end{prf}

\begin{lem}\label{2pt-2}
The Sato type tau-function~$\tau_{\mathrm{s}}$ satisfies the relations
\beq
\e^2\frac{\p^2 \log\tau_{\mathrm{s}}}{\p T_\lambda \p T_\mu} - \frac{1}{p+q}\frac{\lambda\mu}{\lambda+\mu}c_\lambda c_\mu e^{(\lambda+\mu)u/p} \, \in \, \e \A_u[[\e]],\quad \forall\,\lambda,\mu\in\I.
\eeq
\end{lem}

\begin{prf}
Consider the case $\lambda=kp,\mu=lp$ with $k,l\geq1$. 
From the formulae \eqref{leading-coef-L1}, \eqref{leading-coef-L2} and \eqref{f-recursion}, 
it follows that the differential polynomials $f_{kp,l}$ must satisfy
\[
f_{kp,l} - g_{kp,l}e^{(k+l)u} \in \e\A_u[[\e]],
\]
where the constants $g_{kp,l}$ satisfy the recursion relation
\[ 
g_{kp,l}=-\binom{-kp/r}{k+l}-\sum_{j=1}^{l-1}\binom{jp/r}{l-j}g_{kp,j},\quad l\geq1.
\]
Solving this recursion relation we obtain
\beq
g_{kp,l}=-\frac{p}{p+q}\frac{k}{k+l} c_{kp} c_{lp}.
\eeq
Here the numbers $c_\mu$, $\mu\in \I$ are defined in~\eqref{constant-0}. 
Therefore, from~\eqref{dtphi-1} we have
\beq\label{lem-eq-11}
\e\frac{\p\log\phi_1}{\p T_{kp}}=-\sum_{l\geq1}\frac{p}{p+q}\frac{k}{k+l} c_{kp} c_{lp}e^{(k+l)u}z^{-l}+ G_k
\eeq
for some $G_k=G_k(u,u_x,\dots;\e;z)\in\e\A_u[[\e]][[z^{-1}]]$ satisfying $\deg G_k=0$. 

On another hand, from the definition of the Sato type tau-function, 
we know that the derivatives of $\log\tau_{\mathrm{s}}$ with respect to $T's$ are Laurent series in~$\e$ 
with degree being greater or equal to~$-2$. 
So from~\eqref{tau2} it follows that 
\[
\e\frac{\p\log\phi_1}{\p T_{kp}}
= \e \left(e^{-\sum_{l\geq1}\frac{1}{l z^l}\e\frac{\p}{\p T_{lp}}}-1\right)
\frac{\p\log\tau_{\mathrm{s}}}{\p T_{kp}}=-\sum_{l\geq1}\frac{\e^2}{l}\frac{\p^2\log\tau_{\mathrm{s}}}{\p T_{kp}\p T_{lp}} z^{-l}
+\widetilde{G}_k
\]
for some $\widetilde{G}_k=\widetilde{G}_k(u,u_x,\dots;\e;z)\in\e\A_u[[\e]][[z^{-1}]]$ satisfying $\deg \widetilde{G}_k=0$. 
By comparing this formula with~\eqref{lem-eq-11} we obtain 
\[
\e^2\frac{\p^2\log\tau_{\mathrm{s}}}{\p T_{kp}\p T_{lp}}-\frac{p}{p+q}\frac{kl}{k+l} c_{kp} c_{lp}e^{(k+l)u}
\in\e\A_u[[\e]].
\]
Similarly, we can prove the cases when $\lambda$ and $\mu$ take other values. 
The lemma is proved.
\end{prf}

\begin{prp}\label{equivdzs}
The function $\tau_{\mathrm{s}}\left(x+\frac\e{2p}, T; \e\right)$ 
satisfies the defining relations \eqref{taudef1}--\eqref{taudef3} for the tau-function of the FVH.
\end{prp}
\begin{prf}
From the equation~\eqref{Ldecom} it follows that
\[
u(x, T;\e)=\left(1-\lm_1^{-1}\right)\log a_{2,0}(x, T;\e).
\]
Therefore, by using the formula~\eqref{tau-a} we know that
$\tau_{\mathrm{s}}\left(x+ \frac\e{2p},  T; \e\right)$ satisfies
 the equation~\eqref{taudef1}, i.e.
\begin{align}
u(x, T;\e)=\left(\lm_3^{1/2}-\lm_3^{-1/2}\right)\left(\lm_1^{1/2}-\lm_1^{-1/2}\right)
\log\tau_{\mathrm{s}}\left(x+ \frac\e{2p}, T; \e\right).\label{utau}
\end{align}
Let us note that the FVH~\eqref{FVH-0} can be represented in the form
\beq\label{FVH-equiv}
\frac{\p u}{\p T_\mu}=\e^{-1}\left(\lm_1^{1/2}-\lm_1^{-1/2}\right) \lm_1^{-1/2}\res_{\Lambda_3}L^{\mu h}.
\eeq
So by differentiating both sides of~\eqref{utau} with respect to~$T_\mu$ and $T_\lambda$, and by using~\eqref{orproperty} we obtain
\begin{align*}
&\left(\lm_3^{1/2}-\lm_3^{-1/2}\right)\left(\lm_1^{1/2}-\lm_1^{-1/2}\right)
\Omega_{\lambda,\mu}\\
&\qquad
=\e^2\left(\lm_3^{1/2}-\lm_3^{-1/2}\right) \left(\lm_1^{1/2}-\lm_1^{-1/2}\right)
\frac{\p^2 \log\tau_{\mathrm{s}}(x+\frac\e{2p}, T;\e)}{\p T_\lambda \p T_\mu},
\quad \lambda,\mu\in \I.
\end{align*}
Combining with Lemma~\ref{2pt-1} we find that
\[
\Delta_{\lambda,\mu}:=\e^2\frac{\p^2 \log\tau_{\mathrm{s}}(x+\frac\e{2p}, T;\e)}{\p T_\lambda \p T_\mu}
-\Omega_{\lambda,\mu}\in {\rm Ker}\left(\p_x^2\colon \A_u[[\e]]\to \A_u[[\e]]\right).
\]
Thus it follows from 
Lemmas~\ref{leading-omega}, \ref{2pt-1}, \ref{2pt-2} that $\deg \Delta_{\lambda,\mu}=0$ and $\Delta_{\lambda,\mu}\in\e\A_u[[\e]]$, 
so
$\Delta_{\lambda,\mu}$ must be zero.
The proposition is then proved by using Lemma~\ref{dxdt-tau}.
\end{prf}

\section{The topological tau-function of the FVH}\label{topologicaltaufunction} 
In this section, we restrict our discussions to the {\it rational} case, i.e.
\beq\label{zh-5}
p=\frac{1}{m},\quad q=\frac{1}{n},\quad r=-\frac{1}{h},
\eeq
where $m,n$ are coprime positive integers, and $h=m+n$. 

In our previous work \cite{LYZZ}, we showed that the formal power series $Z(x,T; \e)$ 
defined in~\eqref{defZ}
satisfies the following Virasoro constraints:
\begin{equation}\label{vir-zh-di}
L_k Z(x, T; \e)=0,\quad k\ge 0,
\end{equation}
where 
\begin{align}
L_0
=&\sum_{\mu\in\I}\mu T_\mu \frac{\p}{\p T_\mu}
+\frac{x^2}{2m n h\e^2}+\frac{\sigma_1}{24}
-\frac{\Gamma(m)\Gamma(n)}{\Gamma(1+h)}\frac{\p}{\p T_1}, \label{L0op}\\
L_k
=&\sum_{\mu\in\I}\mu T_\mu \frac{\p}{\p T_{\mu+k}}
+\frac x{mn}\frac{\p}{\p T_k} - \frac{\Gamma(m)\Gamma(n)}{\Gamma(1+h)}\frac{\p}{\p T_{1+k}} \label{zh-14} \\
&+\frac{\e^2}{2m}\sum_{\substack{\lambda+\mu=k\\ \lambda,\mu\in \I_1}}
\frac{\p^2}{\p T_\lambda\p T_\mu}
+\frac{\e^2}{2n}\sum_{\substack{\lambda+\mu=k\\ \lambda,\mu\in \I_2}}
\frac{\p^2}{\p T_\lambda\p T_\mu},\quad k\ge 1, \nn
\end{align}
and $\sigma_1$ is defined in \eqref{sigma1sigma3mnh}. From the definitions \eqref{cubicZ}, \eqref{defZ} we know that $Z(x, T; \e)$ has the genus expansion 
\begin{align}
\log Z(x, T;\e) = \frac{1}{mnh\e^2}A + \sum_{g\geq0} (mnh)^{g-1}\e^{2g-2} \widetilde{\mathcal H}_g(t(x, T)), \label{genusexpansionZH}
\end{align}
where $A$ and $\widetilde{\mathcal H}_g(t(x, T))$, $g\geq 0$ are defined in \eqref{def-a} and \eqref{wHHg} respectively.
From~\cite{DLYZ-1} we know that the genus zero part $\widetilde{\mathcal H}_0(t(x, T))$ 
corresponds to a particular solution to the dispersionless FVH, and that  
 $\widetilde{\mathcal H}_g(t(x, T))$, $g\geq 1$ satisfy certain technical conditions.
Based on this we proved in~\cite{LYZZ} that the Virasoro constraints 
given in \eqref{vir-zh-di} lead to the so-called loop equation, which provides a way of determining
$\widetilde{\mathcal H}_g(t(x, T))$, $g\geq 1$ uniquely. 

In order to prove Theorem~\ref{main-1},  we are to construct in this section 
a particular tau-function $\tau_{\mathrm{top}}$ of the FVH which has the genus expansion 
and satisfies the same technical conditions, and to prove in the next sections that it satisfies the Virasoro 
constraints $L_k\tau_{\mathrm{top}}=0$, $k\ge 0$. 
Then Theorem~\ref{main-1} in the rational case follows from the uniqueness theorem of the loop equation 
that was proved in~\cite{LYZZ}. The theorem for the general case will be proved from the one for the rational case by using a continuation argument.

The results of this section are given in the following theorem.

\begin{thm}\label{existencetautopo}
For the rational case, there exists a nontrivial
tau-function 
\[\tau_{\mathrm{top}}=\tau_{\mathrm{top}}(x,T;\e)\in \CC((\e^2))[[x-1, T]]\]
 of the FVH~\eqref{FVH-0} satisfying the following two equations:
\begin{align}
& L_0 \tau_{\mathrm{top}} =0, \label{stringtype}\\
& K_0 \tau_{\mathrm{top}} + \frac{\tau_{\mathrm{top}}}{24} = 0, \label{dilaton}
\end{align}
where $L_0$ is defined in~\eqref{L0op}, and $K_0$ is a linear operator given by 
\begin{align}
K_0:=\sum_{\mu\in \I} T_\mu \frac{\p}{\p T_\mu} 
+ x \frac{\p}{\p x} + \e \frac{\p}{\p \e} 
- \frac{\Gamma(m)\Gamma(n)}{\Gamma(1+h)}  \frac{\p}{\p T_1}. \label{K0op}
\end{align}
Moreover, this tau-function possesses the genus expansion
\beq
\log \tau_{\rm top}(x, T;\e) = \sum_{g\geq0}\e^{2g-2}\F_g(x,T)  \label{zh-15} 
\eeq
for some $\F_g(x,T)\in\CC[[x-1,T]]$, $g\geq 0$. 
Furthermore, for $g\geq 1$,
there exist functions $F_g=F_g(z_0,\dots,z_{3g-2})$ 
such that
\beq\label{tech3}
\F_g(x,T) =F_g\left(v_{\rm top}, v_{\rm top}', \dots, v_{\rm top}^{(3g-2)}\right)
\eeq
and
\begin{align}
& F_1=\frac{1}{24}\log z_1-\frac{mh+n^2}{24nh} z_0 ,  \label{tech1} \\
& \sum_{i\geq1}i z_i \frac{\p F_g}{\p z_i}=(2g-2)F_g, \quad F_g\in z_1^{-(4g-4)}\CC[z_1,\dots,z_{3g-2}], \quad g\geq 2.  \label{tech2} 
\end{align}
Here $v_{\rm top}:=\p_x^2 \F_0$ and ${}'=\p_x$.
\end{thm}

We note that the non-trivial tau-function~$\tau_{\mathrm{top}}$ in the above theorem
is {\it uniquely} determined by \eqref{stringtype}--\eqref{dilaton} 
up to a constant factor 
that is irrelevant for the current study. Thus the identity~\eqref{tech3} for $g=1$ should be 
understood as under the suitable choice of the constant factor. In other words, 
we use the $g=1$ identity of~\eqref{tech3} to determine the constant for~$\tau_{\mathrm{top}}$.
We call this~$\tau_{\mathrm{top}}$
the {\it topological tau-function} of the FVH.
Our construction of~$\tau_{\rm top}$ will consist of five steps:

\noindent \textbf{Step I.} To fix the particular solution $u_{\rm top}(x, T; \e)$ to the FVH, corresponding to $\tau_{\rm top}$, 
by using the initial value problem for the FVH which is implied by the Virasoro constraint $L_0\tau_{\rm top}=0$.

\noindent \textbf{Step II.} To show that the solution $u_{\rm top}(x, T; \e)$ has the following form of genus expansion:
\begin{equation}\label{zh-9}
u_{\rm top}(x, T; \e)=v_{\rm top}+\sum_{g \ge 1}\e^{2g} A_g\left(v_{\rm top}', v_{\rm top}'', \dots, v_{\rm top}^{(3g)}\right),
\end{equation}
where $v_{\rm top}(x, T)=u_{\rm top}(x, T; 0)$, $v'_{\rm top}=\p_x v_{\rm top}$, $v^{(k)}_{\rm top}=\p_x^k v_{\rm top}$.

\noindent \textbf{Step III.} To show a certain transcendency of the function $v_{\rm top}$.

\noindent \textbf{Step IV.} To show the existence of functions $\widetilde{F}_g(z_0, z_1, \dots, z_{3g-2})$ such that
\beq\label{zh-12}
A_g\left(v_{\rm top}', v_{\rm top}'', \dots, v_{\rm top}^{(3g)}\right)=
\p_x^2 \widetilde{F}_g\left(v_{\rm top},v_{\rm top}', v_{\rm top}'', \dots, v_{\rm top}^{(3g-2)}\right), \quad g\ge 1.
\eeq

\noindent \textbf{Step V.} To construct the tau function $\tau_{\rm top}$ satisfying the 
conditions \eqref{taudef1}--\eqref{taudef3}, and show that $\log \tau_{\rm top}$
has the genus expansion of the following form:
\begin{equation}
\log \tau_{\rm top}=\frac{1}{\e^2}\mathcal{F}_0(x,T)+\sum_{g\ge 1}\e^{2g-2}F_g\left(v_{\rm top}, v_{\rm top}', \dots, v_{\rm top}^{(3g-2)}\right),\label{jet-expansion-2}
\end{equation}
where $\mathcal{F}_0(x,T)$ satisfies the relation $v_{\rm top}(x, T)=\p_x^2 \mathcal{F}_0(x,T)$.

\begin{prfn}{Theorem~\ref{existencetautopo}}  
Let us prove the theorem step by step as  outlined above.

\noindent \textbf {Step I.} 
To find the solution $\up(x, T;\e)$ to the FVH, let us fix its initial value~$\up(x, 0;\e)$ 
by using the condition~\eqref{stringtype}. 
Assuming the existence of such a solution with the associated tau-function~$\tp$, 
we apply the operator $\e (\Lambda_3-1) \Lambda_2^{-1/2}$ to both sides 
of the equation $\frac{L_0\tau_{\mathrm{top}}}{\tau_{\mathrm{top}}}=0$, and put $T=0$, then we arrive at the equation
\begin{equation}\label{zh-7}
\e (\Lambda_3-1)\Lambda_2^{-1/2}\left.\left(\frac{x^2}{2m n h \e^2}
-\frac{\Gamma(m)\Gamma(n)}{\Gamma(1+h)} \frac{\p\log\tp}{\p T_1}\right)\right|_{T=0}=0.
\end{equation}
Using the defining relation \eqref{taudef2} for the tau-functions and the fact that
\[\e(\Lambda_3-1)\Lambda_2^{-1/2}=\e \Bigl(h \e \p_x+\frac{n h}2\e^2 \p_x^2+{\mathcal{O}}\bigl(\e^3\bigr)\Bigr),\]
we obtain from \eqref{zh-7} the following equation:
\begin{equation}\label{zh-4}
\left.\left(\res_{\Lambda_3} L^h \right)\right|_{T=0}=\binom{m+n}{m}\left(x+\frac{n\e}{2}\right).
\end{equation}
It follows from \eqref{zh-1} and~\eqref{zh-5} that
\begin{align}
\res_{\Lambda_3} L^h
=\sum_{0\le \alpha_1\le \dots\le\alpha_m\le n}
\exp\left(\sum_{j=1}^m \up\left(x+\e\alpha_j m-\e (j-1)n, T;\e\right)\right). \label{zh-6}
\end{align}
By using \eqref{zh-4}, \eqref{zh-6} and by making the shift $x\to x-\frac{n}2 \e$ 
we arrive at the equation \eqref{eq-initial} for~$\up(x, 0; \e)$. 

Let us show that there exists a unique solution in $\CC[[x-1;\e]]$ to the equation \eqref{eq-initial}. 
First we observe that the equation \eqref{eq-initial} can be written in the form
\beq\label{equiv117}
	 e^{m V(x;\e)}
	\Biggl(1+\sum_{k\geq1}\e^k\sum_{l\geq1}
	\sum_{\substack{ j_1,\dots,j_l\geq1 \\ 
	j_1+\cdots+j_l=k}} C_{j_1,\dots,j_l} \p_x^{j_1}V(x;\e)\cdots \p_x^{j_l}V(x;\e)
	\Biggr)= x,
\eeq
where $C_{j_1,\dots,j_l}$ are constants depending on~$m,n$. Then 
by substituting $V(x;\e)=\sum_{k\geq0}\e^k V^{[k]}(x)$ into~\eqref{equiv117}, we find that 
$V^{[0]}(x)=\frac1m \log x$, and that $V^{[k]}(x)$, $k\geq 1$ can be uniquely solved in a recursive way. Moreover,
$V^{[k]}(x)\in\CC[[x-1]]$ for $k\geq 0$.
This shows the existence and uniqueness of solution in $\CC[[x-1;\e]]$ to the equation \eqref{eq-initial}.
Combining this with the statement that the FVH~\eqref{FVH-0} is {\it integrable}~\cite{LZZ}, 
we arrive at the following proposition.
\begin{prp}
There is a unique solution 
\[u_{\rm top}=u_{\rm top}(x, T;\e)\in \CC[[x-1, T;\e]]\]
to the FVH satisfying the initial condition
\beq\label{iniproblem}
u_{\rm top}(x, 0;\e) = V(x;\e),
\eeq
where $V(x;\e)$ is the unique solution in $\CC[[x-1;\e]]$ to the equation \eqref{eq-initial}.
We call $u_{\rm top}$ the topological solution to the FVH. 
\end{prp}

We show next that $V(x; \e)$ and $u_{\rm{top}}(x, T; \e)$ can be represented as power series of $\e^2$. 
To see this let us introduce some notations. Let $\YY$ denote the set of partitions of non-negative integers, i.e., 
\[
\YY:= \biggl\{(J_1,J_2,\dots)\in\mathbb{Z}^\mathbb{N}\mid J_1\geq J_2 \geq \cdots \geq0, \ \sum_{i\ge 1} J_i<\infty \biggr\}.
\]
For a partition $J=(J_1,J_2,\dots)\in\YY$, denote by~$ l(J)$ the number of non-zero 
components of~$J$, by $|J|:= J_1+ J_2 +\dots$ the weight of~$J$, and by $\YY_k$
the set of all partitions of weight~$k$. We have the following lemma.
\begin{lem}\label{lem-coef-of-fvh}
For each $\lambda\in\I$, we have
\begin{equation}
\lm_1^{-\frac12}\res_{\Lambda_3} L^{\lambda h}
=c_\lambda e^{\lambda m u}\sum_{g\geq0}\e^{2g} M_\lambda^{[g]}\left(u',\cdots,u^{(2g)}\right), \label{coef-of-fvh}
\end{equation}
where $c_\lambda$ is defined in~\eqref{constant-0},
\begin{equation}
M_\lambda^{[g]}:=M_\lambda^{[g]}\left(u',\cdots,u^{(2g)}\right)
=\sum_{J\in\YY_{2g}} a_J (\lambda) u^{(J)} \in \mathcal{A}_u, ~g\geq 0,
\label{M-expr}
\end{equation}
and 
\[u^{(k)}:=\p_x^k u, \quad  
u^{(J)}:= u^{(J_1)}\cdots u^{(J_{ l(J)})},  \quad k\geq0, \ J=(J_1,J_2,\dots).\]
Moreover, the coefficient $a_{\emptyset}(\lambda)=1$, and
the coefficients $a_{J} (\lambda)$, $J\in \YY_{\rm even}$ are polynomials of~$\lambda$
with degrees less than or equal to $ l(J)+|J|/2$. 
\end{lem}
\begin{prf}
We first show that 
$\lm_1^{-\frac12}\res_{\Lambda_3} L^{\lambda h}\in\mathcal{A}_u[[\e^2]]$ for each~$\lambda\in \I$.
Indeed, 
\begin{align}
		\lm_1^{-\frac12}\res_{\Lambda_3} L^{\lambda h}
		&=\lm_1^{-\frac12}\res_{\Lambda_3} \left(\lm_2^{-1}+e^{\lm_1u} \lm_1\right)^{\lambda h} \label{equality1}\\
		&=\lm_1^{\frac12}\res_{\Lambda_3} \left(\lm_2^{-1}+e^u \lm_1\right)^{\lambda h}, \label{equality2}
\end{align}
which shows that 
$\lm_1^{-\frac12}\res_{\Lambda_3} L^{\lambda h}$ is invariant under the map $\e\mapsto -\e$. 
Here, \eqref{equality1} is valid due to the fact that 
\[\res_{\Lambda_3}\sum_{i\in\ZZ}\alpha_i \circ \lm_3^i=\res_{\Lambda_3}\sum_{i\in\ZZ}\lm_3^{-i}\circ\alpha_i,\] 
and \eqref{equality2} is valid because of the fact that 
\[\res_{\Lambda_3}L^{\lambda h}\big|_{u \, \mapsto\, \lm_1 u}=\lm_1\res_{\Lambda_3} L^{\lambda h}.\]

For $\lambda=1/m$, we have
\beq\label{explicitMm}
\lm_1^{-1/2}\res_{\Lambda_3} L^{h/m}=\frac{\lm_3^{1/2}-\lm_3^{-1/2}}{\lm_2^{1/2}-\lm_2^{-1/2}}e^u
=\frac{h}{m}e^u\biggl(1+\sum_{g\geq1}\e^{2g}M_{1/m}^{[g]}\biggr), 
\eeq
from which it follows that $M_{1/m}^{[g]}\in \CC\left[u',\dots,u^{(2g)}\right]$ 
have the form~\eqref{M-expr}. 
Note that Proposition~\ref{taustructure} implies the following tau-symmetry property~\cite{DLYZ-1, DZ-norm}:
\begin{align}
		\frac\p{\p T_{1/m}}
		\left(\lm_1^{-\frac12}\res_{\Lambda_3} L^{\lambda h}\right)
		=\frac\p{\p T_\lambda}
		\left(\lm_1^{-\frac12}\res_{\Lambda_3} L^{h/m}\right), \quad \forall \lambda \in \I. \label{tau-symmetry}
\end{align}
From this equation we arrive at a certain recursion relation for $M_\lambda^{[g]}$, $g\geq 1$, and we find that $M_\lambda^{[g]}$ has the required form~\eqref{M-expr} with 
the coefficients $a_J (\lambda)$ being polynomials of~$\lambda$
with degree less than or equal to $ l(J)+|J|/2$. 
The lemma is proved.
\end{prf}

\begin{lem}\label{lem-initial-form} 
The unique solution in $\QQ[[x-1;\e]]$ to the difference equation~\eqref{eq-initial} can be represented in the form
\beq\label{top-initial-genus0}
V(x;\e)=\frac1m\log x+\sum_{g\geq1}\e^{2g}\frac{P_g(m, n)}{x^{2g}},
\eeq
where $P_g(m, n)\in \QQ\bigl[m,n,m^{-1}\bigr]$ for $g\geq 1$. 
\end{lem}
\begin{prf}
Note that the equation \eqref{eq-initial} can be written as
\beq\label{Vform2}
\res_{\lm_3}\left(\lm^m+e^{V(x-\frac{n\e}2;\e)}\lm^{-n}\right)^h=\binom{m+n}m x.
\eeq
From Lemma~\ref{lem-coef-of-fvh} it follows that this equation can also be represented in the form
\beq\label{Vform3}
e^{m V}\sum_{g\geq0}\e^{2g}M_1^{[g]}\left(V^{(1)},\dots,V^{(2g)}\right) =x,
\eeq
where the coefficients $M_1^{[g]}$ are obtained recursively by using 
the tau-symmetry condition \eqref{tau-symmetry} with $\lambda=1$. 
Then by substituting the expression \eqref{top-initial-genus0} of $V(x;\e)$ into~\eqref{Vform3} one can determine the coefficients $P_g(m,n)$ 
recursively, and show that they belong to $\QQ[m,n,m^{-1}]$.
The lemma is proved. 
\end{prf}

We will give the explicit expressions of some $P_g(m,n)$ in Section~\ref{GAP}.

From the Lemmas \ref{lem-coef-of-fvh}, \ref{lem-initial-form} 
we know that the right-hand side of \eqref{FVH-equiv} 
is in~$\A_u[[\e^2]]$, and that the power series  
$V(x;\e)$ is in $\CC[[x-1;\e^2]]$, thus we arrive at the following corollary.

\begin{cor}\label{top-even-power}
The topological solution $u_{\rm top}$ is an element of  $\CC[[x-1, T;\e^2]]$.
\end{cor}


\noindent \textbf{Step II.} Let us proceed to prove that $\up(x,T;\e)$ 
has the genus expansion~\eqref{zh-9} by using the so-called Euler--Lagrange equation.

\begin{prp}\label{prp-fvh-el}
The formal power series 
$u:=u_{\rm top}$ satisfies the following equation:
\beq\label{el-equation-full}
  \sum_{\lambda\in \mathcal{I}} \lambda \widetilde T_\lambda  {\rm res}_{\Lambda_3} L^{\lambda h} + \frac{x+\frac{n}2\e}{mn}
 = 0,
\eeq
where $\widetilde T_\lambda=T_\lambda-\frac{\Gamma(m)\Gamma(n)}{\Gamma(1+h)}\delta_{\lambda,1}$.
\end{prp}
\begin{prf}
Define a power series $F(x,T;\e)\in\CC[[x-1,T;\e]]$ by 
\beq\label{definitionF}
F(x,T;\e):=\sum_{\lambda\in\I}\lambda \widetilde T_\lambda \res_{\Lambda_3}L_{\rm top}^{\lambda h}+\frac {x+\frac{n}2\e}{mn},
\eeq
where $L_{\rm top}:=\Lambda_2+ e^{u_{\rm top}}\Lambda_1^{-1}$. 
We are to show $F(x,T;\e)\equiv0$. 
By using \eqref{ooproperty}, \eqref{orproperty}, \eqref{FVH-equiv} and \eqref{coef-of-fvh} we obtain, after a lengthy calculation, that 
\begin{align}\label{el-correlator}
&\frac{\p F(x,T;\e)}{\p T_\mu}
=\sum_{k\geq0}\frac{\p \bigl(\res_{\Lambda_3}L^{\mu h}_{\rm{top}}\bigr)}{\p u^{(k)}} \p_x^k\widehat F(x,T;\e),
\quad \forall\,\mu\in\I. 
\end{align}
Here 
$\widehat F(x,T;\e)$ is a power series in $\CC[[x-1,T;\e]]$ defined by
\beq\label{widehatfdef}
\widehat F(x,T;\e)
:=\e^{-1}\left(1-\lm_1^{-1}\right)F(x,T;\e)
=\sum_{\lambda\in\I}\lambda \widetilde T_\lambda\frac{\p u_{\rm top}}{\p T_\lambda}+\frac1m.
\eeq
By taking $T=0$ in~\eqref{definitionF} we obtain 
\[
F(x, 0;\e)=
-\frac{\Gamma(m)\Gamma(n)}{\Gamma(1+h)}\res_{\Lambda_3} L_{\rm top}^h\Big|_{T=0}+\frac{x+\frac n2\e}{mn}.
\]
Thus $F(x, 0;\e)$ vanishes due to the fact that $u_{\rm top}(x, 0;\e)=V(x;\e)$ satisfies the equation \eqref{Vform2} (or equivalently \eqref{eq-initial}).
So from~\eqref{widehatfdef} it follows that
\beq\label{wf0}
\widehat F(x, 0;\e)=\e^{-1}\left(1-\lm_1^{-1}\right)F(x, 0;\e)=0.
\eeq
Then by mathematical induction we know from \eqref{el-correlator}, \eqref{widehatfdef} that
\beq\label{correlator-0}
\frac{\p^iF}{\p T_{\mu_1}\cdots\p T_{\mu_i}}(x, 0;\e)=0,\quad 
\forall\,i\geq0,\ \mu_1,\dots,\mu_i\in\I,
\eeq
which leads to the vanishing of $F(x,T;\e)$. The proposition is proved.
\end{prf}

We call equation~\eqref{el-equation-full} the Euler--Lagrange equation for the FVH.

Define a power series $v_{\rm top}\in\CC[[x-1,T]]$ by
\beq\label{def-v-top}
v_{\rm top}=v_{\rm top}(x,T):=u_{\rm top}(x,T;\e=0).
\eeq
Then it satisfies the following dispersionless FVH: 
\beq\label{zh-10}
\frac{\p v}{\p T_\lambda} = \lambda m n c_\lambda e^{\lambda m v} \frac{\p v}{\p x}, \quad \lambda\in\I.
\eeq
It follows from Proposition~\ref{prp-fvh-el} that 
$v_{\rm top}$ also satisfies the following equation:
\beq\label{el-equation-0}
\sum_{\lambda\in\I}\lambda c_\lambda \widetilde T_\lambda e^{\lambda m v}+\frac{x}{mn}=0.
\eeq
We call \eqref{el-equation-0} the dispersionless Euler-Lagrange equation for the FVH.

\begin{lem}\label{definitionAg}
There exist functions
\beq
A_g=A_g(z_1,\dots,z_{3g})\in z_1^{-(4g-2)}\CC\left[z_1,z_2,\dots,z_{3g}\right], \quad g\geq 1\label{Ag4g22}
\eeq
satisfying the homogeneity condition 
\begin{equation}
\sum_{i\ge1} i z_i \frac{\p A_g}{\p z_i} = 2g A_g, \quad g\geq 1,\label{A-g-homo}
\end{equation}
such that $u_{\rm top}(x,T;\e)$ can be represented in the form
\begin{equation}
u_{\rm top}(x,T;\e)=v_{\rm top}(x,T)+\sum_{g\geq1}\e^{2g} A_g\left(v_{\rm top}'(x,T),\dots,v_{\rm top}^{(3g)}(x,T)\right). \label{quasi-miura} 
\end{equation}
Here $v_{\rm{top}}^{(k)}(x, T)=\p_x^k v_{\rm{top}}(x, T)$.
\end{lem}
\begin{prf}
According to Corollary \ref{top-even-power}, $u_{\rm top}(x,T;\e)$ can be represented in the form
\beq\label{top-expan}
u_{\rm top}(x,T;\e)=\sum_{g\geq0}\e^{2g} u_{\rm top}^{[g]}(x,T),\quad u_{\rm top}^{[0]}(x,T)=v_{\rm top}(x,T),
\eeq
where $u_{\rm top}^{[g]}(x,T)=:u_{\rm top}^{[g]}\in\CC[[x-1,T]]$. 
The Euler--Lagrange equation \eqref{el-equation-full} for $u=u_{\rm top}$ 
is then equivalent to \eqref{el-equation-0} together with the following recursion relations for 
$u_{\rm top}^{[g]}$, $g\geq 1$:
\begin{align}\label{recursion-u-g}
&u^{[g]}_{\rm top} \sum_{\lambda\in\I}\lambda^2 c_\lambda \widetilde T_\lambda e^{\lambda m v_{\rm top}}\\
=& \sum_{J\in\YY_{\rm even}} \sum_{\lambda\in\I}\lambda c_\lambda 
a_J(\lambda) \widetilde T_\lambda e^{\lambda m v_{\rm top}} \sum_{n\ge\ell(J)}\sum_{\substack{0\leq g_1,\dots,g_n\leq g-1 \\g_1+\dots+g_n =g-\frac{|J|}2}} b_{J;g_1,\dots,g_n}^{[g]}
\prod_{i=1}^n\p_x^{J_i}u_{\rm top}^{[g_i]} ,\notag
\end{align}
where $b_{J;g_1,\dots,g_n}^{[g]}$ are some constants and 
$a_J(\lambda)$ are the coefficients that appear in~\eqref{M-expr}.

Define a sequence of power series (cf. \cite{IZ}) by
\beq
Q_k:= 
\sum_{\lambda\in\I}\lambda^{k+1} c_\lambda \widetilde T_\lambda e^{\lambda m v_{\rm top}},
\quad k\geq1. \label{Qkiz}
\eeq
By taking the $x$-derivative of \eqref{el-equation-0} and~\eqref{Qkiz}  
we find that $Q_k$, $k\geq1$ satisfy the relations
\begin{equation}\label{Q-zh-ch}
Q_1=-\frac1{m^2n v_{\rm top}'},\quad
Q_{k}=\frac1{m v_{\rm top}'}\p_x Q_{k-1},\qquad 
k\geq2.
\end{equation}
So we have
\begin{equation}\label{Q-deg}
Q_k\in\left(v_{\rm top}'\right)^{1-2k}\CC\left[v_{\rm top}',\dots,v_{\rm top}^{(k)}\right],\quad
\sum_{i\geq1} i v_{\rm top}^{(i)}\frac{\p Q_k}{\p v_{\rm top}^{(i)}}=-Q_k,\quad k\geq1.
\end{equation}

When $g=1$, we know from \eqref{recursion-u-g}, \eqref{Q-zh-ch} that $u_{\rm{top}}^{[1]}=A_1(v_{\rm top}',v_{\rm top}'', v_{\rm top}''')$, where
\[A_1(z_1, z_2, z_3)=\frac{nh}{24}\left(\frac{z_3}{z_1}-\frac{z_2^2}{z_1^2}+z_2\right).\]
It is obvious that $A_1$ satisfies the conditions \eqref{Ag4g22}, \eqref{A-g-homo}.
Assume that we have already found the functions
\beq\label{A-zh-g-1}
A_{g'}=A_{g'}(z_1,\dots,z_{3g'})\in z_1^{-(4g'-2)}\CC[z_1,\dots,z_{3g'}], \quad g'\leq g-1
\eeq
satisfying the homogeneity condition 
\beq\label{A-g-degprime}
\sum_{i\geq1}i z_i\frac{\p A_{g'}}{\p z_i}=2g' A_{g'},
\eeq
and that 
$u_{\rm top}^{[g']}:=A_{g'}\Bigl(v_{\rm top}^{(1)},\dots,v_{\rm top}^{(3g')}\Bigr)$ satisfy the equation~\eqref{recursion-u-g} 
for $g'\leq g-1$. 
According to Lemma~\ref{lem-coef-of-fvh}, 
$a_{J}(\lambda)$ are polynomials of~$\lambda$ with degrees less than or equal to 
$ l(J)+|J|/2$, so it follows from~\eqref{Q-deg} that
\beq
\sum_{\lambda\in\I}\lambda c_\lambda \widetilde T_\lambda a_{J}(\lambda)e^{\lambda m v_{\rm top}}
\in \left(v_{\rm top}'\right)^{1-2 l(J)-|J|}\CC\Bigl[v_{\rm top}',\dots, v_{\rm top}^{ \left( l(J)+|J|/2 \right)} \Bigr],
\quad
\forall\, J\in\YY_{\rm even}. \label{jetspace}
\eeq
Thus by using \eqref{Q-zh-ch}--\eqref{jetspace} 
we know the existence of a unique solution $u_{\rm top}^{[g]}$ to the 
equation~\eqref{recursion-u-g} that satisfies the properties required by the lemma. The lemma is proved.
\end{prf}

\noindent \textbf{Step III.} 
Let us show that the formula \eqref{quasi-miura} yields the quasi-trivial transformation of the 
FVH \cite{DZ-norm,LZ}. Namely, for any solution $v=v(x, T)$ to the dispersionless 
FVH~\eqref{zh-10} with $v'\not\equiv 0$, 
\beq\label{zh-11}
u=v+\sum_{g\geq1}\e^{2g}A_g\left(v',\dots,v^{(3g)}\right)
\eeq
gives a solution to the FVH. To this end, we first prove the following lemma on the transcendency
of the solution~$v_{\rm{top}}$ to the dispersionless FVH~\eqref{zh-10}.

\begin{lem}\label{diff-poly-vani}
Let $F\left(z_0,z_1,\dots,z_N\right)\in \A_{z_0,0}[z_1,\dots,z_N]$ be a polynomial in the
indeterminates $z_1, \dots, z_N$ with coefficients depending smoothly on $z_0$, which
satisfies the condition
\beq\label{vanish-cond}
	F\left(v_{\rm top}(x,T), v_{\rm top}'(x,T),\dots,v_{\rm top}^{(N)}(x,T)\right)=0,
\eeq
then $F(z_0,z_1,\dots,z_N)\equiv0$.
\end{lem}
\begin{prf}
From the dispersionless FVH~\eqref{zh-10}, it follows that	
\[
\frac{\p F}{\p T_\lambda}\Bigl(v_{\rm top},v'_{\rm top},\dots,v^{(N)}_{\rm top}\Bigr)
=n c_\lambda \sum_{j=0}^N\frac{\p F}{\p z_j}\Bigl(v_{\rm top},v'_{\rm top},\dots,v^{(N)}_{\rm top}\Bigr) \, 
\p_x^{j+1}\Bigl(e^{\lambda m v_{\rm top}}\Bigr),\quad \lambda\in \I.
\]
Dividing the right-hand side of the above equation 
by $e^{\lambda m v_{\rm top}}$ we obtain a polynomial in~$\lambda$ with degree $N+1$. 
Since this polynomial vanishes for any $\lambda\in\I$, 
all its coefficients must vanish. Thus we obtain
\[
\frac{\p F}{\p z_j}\left(v_{\rm top},v'_{\rm top},\dots, v_{\rm top}^{(N)}\right)=0,\quad j=0,\dots,N.
\]
For a similar reason we find that for all $k\geq 1$, $0\leq j_1,\dots,j_k\leq N$,
\[
\frac{\p^k F}{\p z_{j_1}\cdots \p z_{j_k}}\left(v_{\rm top},v'_{\rm top},\dots, v_{\rm top}^{(N)}\right)=0,
\]
which lead to the vanishing of $F$. The lemma is proved.
\end{prf}

Now let us prove that the function~$u$ defined by~\eqref{zh-11} satisfies the FVH. Indeed, 
for any $\lambda \in\I$, it follows from Theorem 1.2 of \cite{LWZ}
the existence of a quasi-trivial transformation 
\[
u=v+\sum_{g\geq1}\e^{2g}\widehat A_{\lambda,g}\left(v,v',\dots,v^{(N_g)}\right),
\]
which transforms~\eqref{zh-10}
to the $\p_{T_{\lambda}}$-flow of the FVH. 
Here 
\[\widehat A_{\lambda,g}\left(z_0,\dots,z_{N_g}\right)\in z_1^{-N_g'}\A_{z_0,0}\left[z_1,\dots,z_{N_g}\right],\]
and $N_g$ and $N_g'$ are some positive integers.
Since $u_{\rm top}$ satisfies the FVH, from \eqref{quasi-miura} 
it follows that 
\[A_g\left(v_{\rm top}',\dots,v_{\rm top}^{(3g)}\right)
=\widehat A_{\lambda,g} \left(v_{\rm top}, v_{\rm top}',\dots,v_{\rm top}^{(N_g)}\right),\quad g\ge 1.\]
Thus by using Lemma \ref{diff-poly-vani} we arrive at 
\[
A_g\left(z_1,\dots,z_{3g}\right)=\widehat A_{\lambda,g}\left(z_0,z_1,\dots,z_{N_g}\right),\quad \forall\,g\geq1,
\]
so \eqref{zh-11} is the quasi-trivial transformation of the FVH.
	
\noindent\textbf{Step IV.} 
We are to prove the existence of the functions $\widetilde F_g$ satisfying the equations in \eqref{zh-12}.
To this end, we need to use the following lemma which can be proved by applying 
Theorem~1.2 of~\cite{LWZ} to the 
Hamiltonian structure of the FVH given in \cite{LZZ}.
\begin{lem} \label{lem-quasi-trivial}
The quasi-trivial transformation \eqref{zh-12}
transforms the Hamiltonian structure 
\begin{align}
&\frac{\p v}{\p T_\lambda}
= P_0 \left(\frac{\delta H_\lambda^{[0]}}{\delta v(x)}\right),
\label{leading-of-fvh}\\
&P_0=\frac{nh}{m}\p_x,\quad H_\lambda^{[0]}=\frac{c_\lambda}{\lambda h}\int e^{\lambda m v}dx
\end{align}
of the dispersionless FVH  to the one of the FVH given by
\begin{align}
&\frac{\p u}{\p T_\lambda} 
=P \left(\frac{\delta H_\lambda}{\delta u(x)}\right),\\
& P=\frac{\left(\lm_3-1\right)\left(1-\lm_1^{-1}\right)}{\e(\lm_2-1)}, 
\quad  H_\lambda=\frac1{\lambda h}\int \res_{\Lambda_3}L^{\lambda h}dx. \label{Pform}
\end{align}
\end{lem}

It follows from the above Lemma that
\beq\label{PP0}
P\frac{\delta }{\delta u}\int u dx
=P_0 \frac{\delta }{\delta v}\int \biggl(v+\sum_{g\geq1}\e^{2g}A_g\left(v',\dots,v^{(3g)}\right)\biggr)dx,
\eeq
so we have
\[
\frac{\delta}{\delta v}\int A_g\left(v',\dots,v^{(3g)}\right)dx\equiv{\rm const},\quad \forall \, g\geq1. 
\]
By using the homogeneity condition~\eqref{A-g-homo} we know that the constants in the above formula must vanish.
Therefore, there exist functions
\[B_g=B_g(z_0,\dots,z_{3g-1})\in\mathcal{A}_{z_0,0}\left[\log(z_1), z_1,z_1^{-1},z_2,\dots,z_{3g-1}\right],\quad g\ge 1\]
satisfying the relations
\beq\label{ag-bg}
A_g=\sum_{i\geq0}z_{i+1}\frac{\p B_g}{\p z_i},\quad
\sum_{i\geq 1}i z_i\frac{\p B_g}{\p z_i}=(2g-1)B_g.
\eeq
By taking derivatives with respect to~$z_0$ 
on both sides of the above equations we obtain
\beq\label{B1eq}
\sum_{i\geq0}z_{i+1}\frac{\p}{\p z_i}\left(\frac{\p B_g}{\p z_0}\right)=0,  \quad
\sum_{i\geq 1}i z_i\frac{\p }{\p z_i}\left(\frac{\p B_g}{\p z_0}\right)=(2g-1)\frac{\p B_g}{\p z_0}.
\eeq
It follows that $B_g$ does not depend on $z_0$.
Thus by using \eqref{Ag4g22} and \eqref{ag-bg} we arrive at 
\[
B_g\in z_1^{-(4g-3)}\CC[z_1,\dots,z_{3g-1}],\quad g\geq1.
\]

Now let us proceed to prove that $B_g\left(v', v'', \dots, v^{(3g-1)}\right)$
can be represented as the $x$-derivative of a certain function of $v, v', v'',\dots$. We first prove the following lemma.
\begin{lem}\label{zero-cond-1}
If $F\in (v')^{-N'}\A_{v,0}\left[v',\dots,v^{(N)}\right]$ 
satisfies the equations
\begin{align}
& \int  F e^{\lambda m v}v' dx=0,\quad \forall\,\lambda\in\I, \label{zh-13}\\
& \sum_{i\geq1}i v^{(i)} \frac{\p F }{\p v^{(i)}}=N'' F\label{deg-F}
\end{align}
for some positive integers $N,N',N''$, then $F\equiv 0$. 
\end{lem}
\begin{prf}
For any $\lambda\in\I$, a straightforward calculation shows that $F$ satisfies the equation
\beq\label{ns-4}
\frac{\delta }{\delta v}\int F e^{\lambda m v} v' dx
=\sum_{k=0}^N(-1)^k\left(\left(\delta_k F\right)\p_x^k\left(e^{\lambda m v} v'\right)+\left(\p_x^k F\right)\delta_k\left(e^{\lambda m v} v'\right)\right),
\eeq
where  the operators $\delta_k$ are defined by
\[
\delta_k:=\sum_{i\geq0}(-1)^i\binom{i+k}{k}\p_x^i\frac{\p}{\p v^{(i+k)}},\quad k\geq0.
\]
The left-hand side of~\eqref{ns-4} vanishes due to the condition~\eqref{zh-13}. 
Dividing the right-hand side of \eqref{ns-4} by $e^{\lambda m v}$, we obtain 
a polynomial in $\lambda$ of degree $N$. 
The vanishing of this polynomial for any $\lambda\in \I$ yields  
 $\p F/\p v^{(k)}=0$, $k=1,\dots, N$. 
Thus by using the condition~\eqref{deg-F} we conclude that $F\equiv0$. 
The lemma is proved. 
\end{prf}

Taking the derivative with respect to 
$T_\lambda$, $\lambda\in\I$ on both sides of~\eqref{zh-11} 
and integrating with respect to~$x$
we find that
\begin{align}
&\e^{-1}\p_x^{-1}\left(1-\lm_1^{-1}\right)\res_{\Lambda_3} L^{\lambda h}\label{L-v-Bg}\\
=&n c_\lambda e^{\lambda m v}
+n c_\lambda\sum_{i\geq1} 
\p_x^{i+1}\left( e^{\lambda m v}\right)
\frac{\p }{\p v^{(i)}} \sum_{g\geq1}\e^{2g}B_g\left(v',\dots,v^{(3g-1)}\right), \nn
\end{align}
where the FVH~\eqref{FVH-equiv} and its dispersionless limit~\eqref{zh-10} are used. 
Denote 
\[
\Delta_\lambda:=\e^{-1}\p_x^{-1}\left(1-\lm_1^{-1}\right)\res_{\Lambda_3} L^{\lambda h}-n c_\lambda e^{\lambda m v}.
\]
Then from Lemma~\ref{leading-omega} 
one can represent~$\Delta_\lambda$ in the form
\[
\Delta_\lambda=\sum_{k\geq1}\e^{2k}\Delta_\lambda^{[k]}(v,v',\dots).
\]
Here, $\Delta_\lambda^{[k]}\in (v')^{-m_{\lambda, k}}\A_v$ for some nonnegative integers $m_{\lambda, k}$, and $\deg \Delta_\lambda^{[k]}=2k$. 
It follows from the formula~\eqref{orproperty} that 
\[
0=\int \frac{\p \Delta_\lambda}{\p T_\mu} dx
=\int \sum_{k\geq0}\frac{\p \Delta_\lambda}{\p v^{(k)}}\p_x^{k+1}\left(n c_\mu e^{\mu m v}\right)dx
=\mu m n c_\mu
\int e^{\mu m v} v' \left(\frac{\delta}{\delta v}\int \Delta_\lambda dx\right) dx.
\]
Then by employing Lemma~\ref{zero-cond-1} we obtain
\[
\frac{\delta}{\delta v}\int \Delta_\lambda^{[k]} dx=0,\quad k\geq1,
\]
which, together with \eqref{L-v-Bg}, yields the relations 
\beq\label{zero-cond-0}
0=\int \Delta_\lambda dx= \lambda m n c_\lambda\int e^{\lambda m v}v' \left(\frac{\delta}{\delta v}\int \sum_{g\geq1}\e^{2g}B_g\left(v',\dots,v^{(3g-1)}\right) dx\right) dx
\eeq
for all $\lambda\in\I$. By using again Lemma~\ref{zero-cond-1} we obtain
\beq\label{theaboveeqn2}
\frac{\delta}{\delta v} \int B_g\left(v',\dots,v^{(3g-1)}\right) dx=0,\quad \forall\,g\geq1. 
\eeq
Thus by employing~\eqref{ag-bg} and~\eqref{theaboveeqn2} we arrive at the following lemma.
\begin{lem}\label{integration-x}
There exist functions
\begin{align}
&\widetilde F_1(z_0, z_1)=\frac{n h}{24}\log z_1+\frac{n h}{24} z_0, \label{F-1'-expr}\\
&\widetilde F_g(z_1, \dots, z_{3g-2})\in z_1^{-(4g-4)}\CC\left[z_1,\dots,z_{3g-2}\right], \label{F-g'-space}
\quad g\geq2,
\end{align}
satisfying the quasi-homogeneous condition
\beq
\sum_{i\geq1}i z_i\frac{\p \widetilde F_g}{\p z_i}
=(2g-2)\widetilde F_g+\frac{nh}{24}\delta_{g,1}, \quad g\geq1,  \label{F-g'-homo}
\eeq
such that the functions~$A_g$ given in Lemma~\ref{definitionAg} can be represented in the form
\beq
A_g\left(v',\dots, v^{(3g)}\right)
=\p_x^2 \widetilde F_g\left(v,v',\dots,v^{(3g-2)}\right).\label{A-g-int}
\eeq
\end{lem}

\noindent\textbf{Step V.} Finally, let us 
prove the existence of the tau-function $\tp$.
Following the approach of~\cite{Du1}, define $\F_0=\F_0(x,T)\in\CC[[x-1,T]]$ by
\beq\label{def-f-0}
\F_0:=\frac{m n }{2h}\sum_{\lambda,\mu\in\I}
\frac{\lambda \mu}{\lambda+\mu}c_\lambda c_\mu 
\widetilde T_\lambda \widetilde T_\mu e^{(\lambda+\mu)m v_{\rm top}}
+\frac{x}{h}\sum_{\mu\in I} c_\mu \widetilde T_\mu e^{\mu m v_{\rm top}}
+\frac{x^2}{2nh} v_{\rm top}.
\eeq
By using the dispersionless Euler-Lagrange equation~\eqref{el-equation-0} we find that 
$\F_0$ satisfies
\begin{align}
&\p_x^2\F_0=\frac{1}{nh}v_{\rm top},\\
&\p_x \p_{T_\lambda}\F_0= \frac{c_\lambda}h e^{\lambda m v_{\rm top}},\quad \forall\,\lambda\in\I,\label{dxdt-f0} \\
&\p_{T_\lambda} \p_{T_\mu} \F_0=
\frac{mn}{h}\frac{\lambda\mu}{\lambda+\mu}
c_\lambda c_\mu e^{(\lambda+\mu)m v_{\rm top}},\quad \forall\,\lambda,\mu\in\I,\label{dtdt-f0}\\
&\sum_{\lambda\in\I}\lambda \widetilde T_\lambda 
\frac{\p \F_0}{\p T_\lambda}+\frac{x^2}{2mnh}=0,\quad
\sum_{\lambda\in\I}\widetilde T_\lambda 
\frac{\p \F_0}{\p T_\lambda}+x\frac{\p \F_0}{\p x}=2\F_0.\label{string-dilaton-f0}
\end{align}
From Lemmas \ref{definitionAg}, \ref{integration-x} 
we know the existence of functions 
\begin{align}
&F_1=F_1(z_0,z_1):=\frac{1}{24}\log z_1-\frac{mh+n^2}{24nh} z_0,\label{F1-space926}\\
&F_g=F_g(z_0,z_1,\dots,z_{3g-2})\in z_1^{-(4g-4)}\CC[z_1,\dots,z_{3g-2}],\label{Fg-space}
\quad g\geq2,
\end{align}
such that 
\begin{align}
&u_{\rm top}=\left(\lm_3^{1/2}-\lm_3^{-1/2}\right)\left(\lm_1^{1/2}-\lm_1^{-1/2}\right)
\sum_{g\geq0}\e^{2g-2}\F_g,\label{def-f-g}\\
&\sum_{i\geq1}i z_i \frac{\p F_g}{\p z_i}=(2g-2)F_g+\frac{1}{24}\delta_{g,1}, \quad g\geq 1,\label{dilaton-fg}
\end{align}
where 
$\F_g=\F_g(x,T):=F_g\left(v_{\rm top},v_{\rm top}',\dots, v_{\rm top}^{(3g-2)}\right)\in\CC[[x-1,T]]$. 
Clearly, $F_g$ and $\widetilde F_g$ are related by
\beq\label{Fg-tildeFg}
F_g=\frac1{n h}\sum_{k=1}^{g}C_{g-k}(m,n)\p^{2g-2k} \widetilde{F}_k+\frac{C_g(m,n)}{nh}z_{2g-2},\quad
\p:=\sum_{i\geq0}z_{i+1}\frac{\p}{\p z_i},
\eeq
where $C_k(m,n)$, $k\geq 0$ are defined in~\eqref{def-Cg}.

Now let us define a formal power series 
$\tau_{\rm top}\in \CC((\e))[[x-1,T]]$ by the formula
\beq\label{def-tau-top}
\tau_{\rm top}
=\tau_{\rm top}(x,T;\e)
:=e^{\sum_{g\geq0}\e^{2g-2}\F_g}.
\eeq
Then~\eqref{def-f-g} can be written as
\beq\label{u-tau-top}
u_{\rm top}=\left(\lm_3^{1/2}-\lm_3^{-1/2}\right)
\left(\lm_1^{1/2}-\lm_1^{-1/2}\right)\log\tau_{\rm top}.
\eeq
From the identities~\eqref{string-dilaton-f0} it follows that 
\beq\label{vtop0-homo}
\sum_{\lambda\in\I}\lambda \widetilde T_\lambda \frac{\p v_{\rm top}}{\p T_\lambda}+\frac1m=0,\quad
\sum_{\lambda\in\I}\widetilde T_\lambda \frac{\p v_{\rm top}}{\p T_\lambda}+x\frac{\p v_{\rm top}}{\p x}=0,
\eeq
which yield 
\beq\label{jet-degree}
\sum_{\lambda\in\I}\lambda \widetilde T_\lambda \frac{\p v_{\rm top}^{(i)}}{\p T_\lambda}+\frac1m\delta_{i,0}=0,\quad
\sum_{\lambda\in\I}\widetilde T_\lambda \frac{\p v_{\rm top}^{(i)}}{\p T_\lambda}+x\frac{\p v_{\rm top}^{(i)}}{\p x}=-i v_{\rm top}^{(i)}.
\eeq
So by using \eqref{Fg-space} and \eqref{dilaton-fg} we know that $\tp$ satisfies the following
conditions:
\beq\label{top-k0-l0}
L_0\tau_{\rm top}=0,\quad K_0\tau_{\rm top}+\frac1{24}\tau_{\rm top}=0,
\eeq
where the operators $L_0$ and $K_0$ are defined by~\eqref{L0op} and~\eqref{K0op}, respectively.

Now it remains to prove that the power series $\tau_{\rm top}\in\CC((\e))[[x-1,T]]$ is a tau-function of the FVH~\eqref{FVH-0}.
For any given $\lambda,\mu\in\I$, 
 define two power series in $\CC[[x-1,T;\e]]$ by
\begin{align*}
&\Delta_\lambda:=
\left(\lm_3-1\right)\lm_2^{-\frac12}
\e\frac{\p \log\tau_{\rm top}}{\p T_\lambda}
-\res_{\Lambda_3} L^{\lambda h}\big|_{u=u_{\rm top}},\\
&\Delta_{\lambda,\mu}:=
\e^2\frac{\p^2 \log\tau_{\rm top}}{\p T_\lambda\p T_\mu}
-\Omega_{\lambda,\mu}\big|_{u=u_{\rm top}}.
\end{align*}
We are to show that
\beq\label{ns-3}
\Delta_\lambda=0,\quad
\Delta_{\lambda,\mu}=0.
\eeq
Firstly, from the definition~\eqref{def-tau-top} of~$\tau_{\rm top}$, equations \eqref{dxdt-f0}, \eqref{dtdt-f0} and
Lemmas \ref{leading-omega}, \ref{diff-poly-vani} we know the existence of 
\begin{align}
 &\Delta_\lambda^*=\Delta^*_\lambda(z_0,z_1,\dots), \quad\Delta_{\lambda,\mu}^*=\Delta^*_{\lambda,\mu}(z_0,z_1,\dots)\in \e\A_{z_0,0}[z_1^{-1},z_1,z_2,\dots][[\e]] \label{homogeneousDeltaDeltastar}
\end{align}
satisfying 
\[\Delta_\lambda=\Delta^*_\lambda\left(v_{\rm top},v_{\rm top}',\dots\right), \quad 
\Delta_{\lambda,\mu}=\Delta^*_{\lambda,\mu}\left(v_{\rm top},v_{\rm top}',\dots\right).\] 
Then by using Lemmas~\ref{leading-omega}, \ref{diff-poly-vani}, 
the second equality of~\eqref{top-k0-l0} and the second identity of~\eqref{jet-degree} 
we find that
\beq
\left(\sum_{i\geq1}i z_i \frac{\p}{\p z_i}- \e \frac{\p }{\p\e}\right)\Delta_\lambda^*
=0,\quad\left(\sum_{i\geq1}i z_i \frac{\p}{\p z_i}- \e \frac{\p }{\p\e}\right)\Delta_{\lambda,\mu}^*
=0.\label{delta-zero-1}
\eeq
Next, by applying $\e\p_{T_\lambda}$ and $\e^2\p_{T_\lambda}\p_{T_\mu}$ 
to both sides of \eqref{u-tau-top} we obtain
\[
\p_x \Delta_\lambda=0,\quad
\p_x^2 \Delta_{\lambda,\mu}=0.
\]
Then by using Lemma~\ref{diff-poly-vani} we arrive at
\[
\p \Delta^*_\lambda=0,\quad 
\p^2\Delta^*_{\lambda,\mu}=0,
\]
where $\p$ is defined in \eqref{Fg-tildeFg}. Together with the formulae \eqref{homogeneousDeltaDeltastar}--\eqref{delta-zero-1} this implies that
\[
\Delta^*_\lambda=0,\quad 
\Delta^*_{\lambda,\mu}=0.
\]
Hence the identities~\eqref{ns-3} hold true.
It follows that 
the power series $\tau_{\rm top}$ satisfies 
the definition~\eqref{taudef1}--\eqref{taudef3} 
of the tau-function of the FVH. Theorem~\ref{existencetautopo} is proved.
\end{prfn}

\section{Virasoro constraints}\label{zh-8}
The goal of this section is to show, following the approaches used in \cite{ASM, BW, OS},
that the topological tau-function~$\tau_{\rm top}$ of the FVH 
satisfies the Virasoro constraints 
$L_k\tp=0$, $k\geq 0$, where the operators $L_k$ are defined in~\eqref{L0op}, \eqref{zh-14}.

We will first derive some useful formulae valid for arbitrary power series solutions to the FVH. 
Let $u$ be a solution to the FVH in the ring $\CC[[x-1, T;\e]]$, 
$(\Phi_1,\Phi_2,\tau)$  
a dressing triple associated to~$u$, and $\psi_1, \psi_2$ 
the wave functions corresponding to $\Phi_1,\Phi_2$. See their definitions given in Section~\ref{sectiontau}.
Introduce the following notations:

\noindent (i) $\Gamma_1$ and~$\Gamma_2$ denote the following 
elements in $\CC[\lm_3,\lm_3^{-1},\e][[x-1, T]]$:
\begin{align}
&\Gamma_1:=\frac{x}{h\e}\lm_3^{-1}
+\frac{m}{\e}\sum_{\mu\in\J_1}\mu T_\mu\lm_3^{\mu m-1}
+\frac{m}{2\e}\sum_{\mu\in\J_3}\mu T_\mu\lm_3^{\mu m-1},\\
&\Gamma_2:=-\frac{x}{h\e}\lm_3
-\frac{n}{\e}\sum_{\mu\in\J_2}\mu T_\mu \lm_3^{-\mu n+1}
-\frac{n}{2\e}\sum_{\mu\in\J_3}\mu T_\mu \lm_3^{-\mu n+1}.
\end{align}
Here $\J_1, \J_2, \J_3$ are defined in \eqref{zh-18}.

\noindent (ii) $M_1$ and~$M_2$ denote the following difference operators: 
\beq
 M_1:=\Phi_1 \Gamma_1 \Phi_1^{-1},	\quad 	M_2:=\Phi_2 \Gamma_2 \Phi_2^{-1}.
\eeq
Clearly, $M_1\in\CC((\lm_3^{-1},\e))[[x-1, T]]$, $M_2\in\CC((\lm_3,\e))[[x-1, T]]$. 

\noindent (iii) 
$N_1$ and $N_2$ denote the following generating series of difference operators:
\begin{align}
& N_1=N_1(\xi,\zeta) :=\sum_{k\geq1}\frac{(\zeta-\xi)^k}{(k-1)!}
\sum_{d\in \mathbb{Z}}\xi^{-k-d} M_1^{k-1} L^{(k+d-1)h/m},\label{N1}\\
& N_2=N_2(\xi,\zeta) :=\sum_{k\geq1}\frac{(\zeta-\xi)^k}{(k-1)!}
\sum_{d\in \mathbb{Z}}\xi^{-k-d} M_2^{k-1} L^{(k+d-1)h/n} .\label{N2}
\end{align}
We note that for $k\in \ZZ$, $L^{k/m} \in \A_u[[\e]]((\Lambda_3^{-1}))$, $L^{k/n}\in\A_u[[\e]]((\Lambda_3))$.
It is then clear that 
\begin{align*}
&N_1\in\CC((\lm_3^{-1},\e))[[x-1, T]][[\xi,\xi^{-1}]][[\zeta-\xi]],\\ &N_2\in\CC((\lm_3,\e))[[x-1, T]][[\xi,\xi^{-1}]][[\zeta-\xi]].
\end{align*}
See~\cite{LZZ} for the definition of the fractional powers of $L$.

\noindent (iv) $X_1$ and $X_2$ denote the following two operators:
\begin{align}
&X_1(\xi,\zeta)
=e^{\vartheta_1(x, T;\e;\zeta)-\vartheta_1(x, T;\e;\xi)}
\exp\Biggl(\sum_{\mu\in\I_1}
\left(\frac1{\xi^{\mu m}}-\frac1{\zeta^{\mu m}}\right)\frac1{\mu m}\e\frac{\p}{\p T_\mu}\Biggr),\\
&X_2(\xi,\zeta)
=e^{\vartheta_2(x, T;\e;\zeta)-\vartheta_2(x, T;\e;\xi)}
\exp\Biggl(\sum_{\mu\in\I_2}
\left(\frac1{\zeta^{\mu n}}-\frac1{\xi^{\mu n}}\right)\frac1{\mu n}\e\frac{\p}{\p T_\mu}\Biggr),
\end{align}
where $\vartheta_1$, $\vartheta_2$ are defined in \eqref{zh-19}, \eqref{zh-20}.

\noindent (v) Denote by $Y_k$, $k\geq 0$ the following operators in $\CC((\e))[[\lm_3,\lm_3^{-1}]][[x-1,T]]$:
\beq\label{Ym-def}
Y_k:=\frac1m M_1 L^{(k+1/m)h}-\frac1n M_2 L^{(k+1/n)h}
-\frac{\Gamma(m)\Gamma(n)}{\Gamma(1+h)\e}L^{(k+1)h}.
\eeq

\begin{lem}\label{vertexoperator}
The following formulae hold true:
\begin{align}
&N_1(\xi,\zeta)\psi_1(x, T;\e;z) \label{vo1} \\
= & (\zeta-\xi) \sum_{d\in\ZZ}\xi^{-d}z^{d-1} 
\frac{X_1(\xi,\zeta)\tau_{\mathrm{s}}(x, T-[z^{-1}]_1;\e)}{\tau_{\mathrm{s}}(x, T-[z^{-1}]_1;\e)} \psi_1(x, T;\e;z), \nn \\
& N_2(\xi,\zeta)\psi_2(x, T;\e;z) \label{vo2} \\
= & (\zeta-\xi)\sum_{d\in\ZZ}\xi^{-d}z^{d-1}
\frac{X_2(\xi,\zeta)\tau_{\mathrm{s}}(x+h\e, T-[z^{-1}]_2;\e)}{\tau_{\mathrm{s}}(x+h\e, T-[z^{-1}]_2;\e)}
\psi_2(x, T;\e;z). \nn
\end{align}
\end{lem}
\begin{prf}
The generating series $N_1$ can be written as
\[
N_1(\xi,\zeta):=(\zeta-\xi)e^{(\zeta-\xi)M_1}\sum_{d\in\ZZ}\xi^{-d}L^{(d-1)h/m}.
\]
So we have 
\begin{align}
&\frac{1}{\zeta-\xi}N_1(\xi,\zeta)\psi_1(x, T;\e; z)
=\sum_{d\in\ZZ}\xi^{-d}z^{d-1}e^{(\zeta-\xi)\p_z}\psi_1(x, T; \e; z) \nn \\
=&\sum_{d\in\ZZ}\xi^{-d}z^{d-1}e^{\vartheta_1(x, T;\e;z+\zeta-\xi)}
\frac{\tau_{\mathrm{s}}(x, T-[(z+\zeta-\xi)^{-1}]_1;\e)}{\tau_{\mathrm{s}}(x, T;\e)} \nn \\
=&\sum_{d\in\ZZ}\xi^{-d}z^{d-1}e^{\vartheta_1(x, T;\e;z+\zeta-\xi)-\vartheta_1(x, T;\e;z)}
\frac{\tau_{\mathrm{s}}(x, T-[(z+\zeta-\xi)^{-1}]_1;\e)}{\tau_{\mathrm{s}}(x, T-[z^{-1}]_1;\e)}\psi_1(x, T;\e;z) \nn \\
=&\sum_{d\in\ZZ}\xi^{-d}z^{d-1}e^{\vartheta_1(x, T;\e;\zeta)-\vartheta_1(x, T;\e;\xi)}
\frac{\tau_{\mathrm{s}}(x, T-[\zeta^{-1}]_1;\e)}{\tau_{\mathrm{s}}(x, T-[\xi^{-1}]_1;\e)}\psi_1(x, T;\e;z) \nn \\
=&\sum_{d\in\ZZ}\xi^{-d}z^{d-1}e^{\vartheta_1(x, T;\e;\zeta)-\vartheta_1(x, T;\e;\xi)}
\frac{\tau_{\mathrm{s}}(x, T-[z^{-1}]_1-[\zeta^{-1}]_1+[\xi^{-1}]_1;\e)}{\tau_{\mathrm{s}}(x, T-[z^{-1}]_1;\e)}\psi_1(x, T;\e;z) \nn\\
=&\sum_{d\in\ZZ}\xi^{-d}z^{d-1}\frac{X_1(\xi,\zeta)\tau_{\mathrm{s}}(x, T-[z^{-1}]_1;\e)}{\tau_{\mathrm{s}}(x,T-[z^{-1}]_1;\e)}
\psi_1(x, T;\e;z). \nn
\end{align}
Here, the following property of the delta-function is used:
\[
f(z)\sum_{d\in\ZZ}\xi^{-d}z^{d-1}=f(\xi)\sum_{d\in\ZZ}\xi^{-d}z^{d-1},
\quad \forall \, f(z)\in R[[z,z^{-1}]],
\]
where $R$ is any given ring.  So the formula \eqref{vo1} holds true. The formula~\eqref{vo2} can be proved in a similar way. 
The lemma is proved. 
\end{prf}

\begin{lem}\label{lem-yl}
We have the following relations between the operators $Y_k$ and the Virasoro operators~$L_k$, $k\geq 0$:
\begin{align}
&\frac{-(Y_k)_-\psi_1}{\psi_1}
=\left(e^{\sum_{\mu\in\I_1}\frac {-z^{-\mu m}}{\mu m}\e\frac{\p}{\p T_\mu}}-1\right)\frac{\widetilde L_k\tau_{\mathrm{s}}}{\tau_{\mathrm{s}}},\label{Ypsi1}\\
&\frac{(Y_k)_+\psi_2}{\psi_2}=\left(e^{h\e\p_x+\sum_{\mu\in\I_2}\frac {z^{-\mu n}}{\mu n}\e\frac{\p}{\p T_\mu}}-1\right)\frac{\widetilde L_k \tau_{\mathrm{s}}}{\tau_{\mathrm{s}}},\label{Ypsi2}
\end{align}
where 
\beq\label{def-tildeL}
\widetilde L_k:=L_k-\frac{x}{2mn\e}\delta_{k,0}-\frac{h\e}{2mn}\frac{\p}{\p T_k}, \quad k\ge 0.
\eeq
\end{lem}

\begin{prf}
From the definition~\eqref{Ym-def} of~$Y_k$ and 
the definitions~\eqref{N1}, \eqref{N2} of $N_1,N_2$ we obtain
\begin{align}
-(Y_k)_-
=&
-\frac1{2m}\res_\xi \xi^{k m+1}\left(\p_{\zeta}^2 N_1(\xi,\zeta)|_{\zeta=\xi}\right)_-
+\frac1{2n}\res_\xi \xi^{k n+1}\left(\p_{\zeta}^2 N_2(\xi,\zeta)|_{\zeta=\xi}\right)_-   \label{Yk-N1-N2} \\
&+\frac{\Gamma(m)\Gamma(n)}{\Gamma(1+h)\e}\left(L^{(k+1)h}\right)_-. \nn
\end{align}
To show~\eqref{Ypsi1}, 
we need to represent  
$-\frac1{\psi_1}\bigl(L^{k h}\bigr)_-\psi_1$, $-\frac1{\psi_1}(N_i)_-\psi_1$, $i=1,2$ in terms of~$\tau_{\mathrm s}$.

From \eqref{defwavefunctions}, \eqref{phiPhi1} and~\eqref{tau2} 
it follows that
\beq\label{Lpsi-tau}
\frac{-\left(L^{k h}\right)_-\psi_1}{\psi_1}
=\left(e^{-\sum_{\mu\in\I_1}\frac {z^{-\mu m}}{\mu m}\e\frac{\p}{\p T_\mu}}-1\right)
\frac{\e\p_{T_k}\tau_{\mathrm{s}}}{\tau_{\mathrm{s}}},\quad k\geq1.\\
\eeq
To  calculate $-(N_1)_-\psi_1/\psi_1$ and $-(N_2)_-\psi_1/\psi_1$, we
observe that $N_1$ and $N_2$ can be decomposed into the following forms:
\beq\label{N1-N2-decom}
N_1=f_1\left(L^{h/m}\right)+g_1\left(L^{h/n}\right), \quad
N_2=f_2\left(L^{h/m}\right)+g_2\left(L^{h/n}\right),
\eeq
where $f_1(z)$, $f_2(z)$, $g_1(z)$ and $g_2(z)$ are defined by
\begin{align}
&f_1(z):=-\frac{X_1(\xi,\zeta)\tau_{\mathrm{s}}(x, T-[z^{-1}]_1;\e)}
{\tau_{\mathrm{s}}(x, T-[z^{-1}]_1;\e)}
\frac{1-\zeta/z}{1-\xi/z}, \label{deff1}\\
&g_1(z):=\frac{\zeta}{\xi}\frac{X_1(\xi,\zeta)\tau_{\mathrm{s}}(x+h\e, T-[z^{-1}]_2;\e)}
{\tau_{\mathrm{s}}(x+h\e, T-[z^{-1}]_2)},\label{defg1}\\
&f_2(z):=\frac{\zeta}{\xi}\frac{X_2(\xi,\zeta)\tau_{\mathrm{s}}(x, T-[z^{-1}]_1;\e)}
{\tau(x, T-[z^{-1}]_1;\e)}, \label{deff2}\\
&g_2(z):=-\frac{X_2(\xi,\zeta)\tau_{\mathrm{s}}(x+h\e, T-[z^{-1}]_2;\e)}
{\tau_{\mathrm{s}}(z+h\e, T-[z^{-1}]_2;\e)}
\frac{1-\zeta/z}{1-\xi/z}. \label{defg2}
\end{align}
Indeed, denote 
$$
D_0:=e^{\frac{1}{2\e}\sum_{k\geq1}T_k L^{k h}}
\in\CC[\lm_3,\lm_3^{-1}][[ T;\e^{-1}]].
$$
Then by applying~$D_0$ on both sides of the bilinear equations~\eqref{bilinearid12}, we obtain  
\begin{align}
& \res_z z^{\frac{x}{h\e}}
e^{\frac1\e\sum_{\mu\in\I_1}T_\mu z^{\mu m}}
\frac{\tau_{\mathrm{s}}(x, T-[z^{-1}]_1;\e)}{\tau_{\mathrm{s}}(x, T;\e)}
\psi_1^*(x+\ell h\e, T';\e;z)\frac{dz}{z}\label{bi-1}\\
 =& \res_z z^{-\frac{x}{h\e}}
e^{\frac1\e\sum_{\mu\in J_2}T_\mu z^{\mu n}}
\frac{\tau_{\mathrm{s}}(x+h\e, T-[z^{-1}]_2;\e)}{\tau_{\mathrm{s}}(x, T;\e)}
\psi_2^*(x+\ell h\e, T';\e;z)\frac{dz}{z}, \quad \forall\,\ell\in\ZZ. \nn 
\end{align} 
By applying the operator
\[
\frac{\tau_{\mathrm{s}}\left(x, T-[\zeta^{-1}]_1+\left[\xi^{-1}\right]_1;\e\right)}{\tau_{\mathrm{s}}(x, T;\e)}
X_1(\xi,\zeta)
\]
on both sides of~\eqref{bi-1}
we arrive at the identities
\begin{align}
&\res_z 
\frac{X_1(\xi,\zeta)\tau_{\mathrm{s}}\left(x, T-[z^{-1}]_1;\e\right)}{\tau_{\mathrm{s}}\left(x, T-[z^{-1}]_1;\e\right)}
\frac{1-z/\zeta}{1-z/\xi}
\bigl(D_0\psi_1(x, T;\e;z)\bigr)
\psi_1^*(x+\ell h\e, T';\e;z)\frac{dz}{z} \label{eq-1}\\
 = & \res_z
\frac{X_1(\xi,\zeta)\tau_{\mathrm{s}}\left(x+h\e, T-[z^{-1}]_2;\e\right)}{\tau_{\mathrm{s}}\left(x+h\e, T-[z^{-1}]_2\right)}
\bigl(D_0\psi_2(x, T;\e;z)\bigr) 
\psi_2^*(x+\ell h\e, T';\e;z)\frac{dz}{z}. \nn
\end{align}
On another hand, by using the fact that 
\[
(\zeta-\xi)\sum_{d\in\ZZ}\xi^{-d}z^{d-1}=-\frac{1-\zeta/z}{1-\xi/z}
+\frac{\zeta}{\xi}\frac{1-z/\zeta}{1-z/\xi},
\]
we can rewrite equation~\eqref{vo1} in the form
\[
N_1\psi_1=f_1(z)\psi_1
+\frac\zeta\xi\frac{X_1(\xi,\zeta)\tau_{\mathrm{s}}(x, T-[z^{-1}]_1;\e)}
{\tau_{\mathrm{s}}(x, T-[z^{-1}]_1;\e)}
\frac{1-z/\zeta}{1-z/\xi}\psi_1.
\]
So it follows from \eqref{eq-1} and the equations $L^{h/m}\psi_1=z\psi_1$,  $L^{h/n}\psi_2=z\psi_2$ that
\begin{align*}
&\res_z
\left[\left(N_1-f_1(L^{h/m})\right)D_0\psi_1(x, T;\e;z)\right]
\psi_1^*(x+\ell h\e, T';\e;z)\frac{dz}{z}\\
=&\res_z
\left[g_1(L^{h/n})D_0\psi_2(x, T;\e;z)\right]
\psi_2^*(x+\ell h\e, T';\e;z)\frac{dz}{z}, \quad \forall\,\ell\in \ZZ.
\end{align*}
From the above equations and Lemma~\ref{d1equaltod2} we obtain the first identity of~\eqref{N1-N2-decom}.
The proof for the second identity of~\eqref{N1-N2-decom} is similar. 

By using \eqref{N1-N2-decom} and the identities 
\[
\left[g_1\left(L^{h/n}\right)\right]_-=0,\quad \left[g_2\left(L^{h/n}\right)\right]_-=0
\]
we find that
\[
(N_1)_-=f_1\left(L^{h/m}\right)-\res_{\Lambda_3} f_1\left(L^{h/m}\right),
\quad
(N_2)_-=f_2\left(L^{h/m}\right)-\res_{\Lambda_3}f_2\left(L^{h/m}\right).
\]
Together with~\eqref{deff1}--\eqref{defg2} we obtain
\begin{align}
& -\frac{\left(N_1\right)_-\psi_1}{\psi_1}
=\left(e^{-\sum_{\mu\in\I_1}\frac {z^{-\mu m}}{\mu m}\e\frac{\p}{\p T_\mu}}-1\right)
\left(\frac{X_1(\xi,\zeta)\tau_{\mathrm{s}}(x,T;\e)}{\tau_{\mathrm{s}}(x,T;\e)}+\widetilde N_1\right),\label{N1psi} \\
& - \frac{\left(N_2\right)_-\psi_1}{\psi_1}
= - \left(e^{-\sum_{\mu\in\I_1}\frac {z^{-\mu m}}{\mu m}\e\frac{\p}{\p T_\mu}}-1\right)
\left(\frac{\zeta}{\xi}\frac{X_2(\xi,\zeta)\tau_{\mathrm{s}}(x, T;\e)}{\tau_{\mathrm{s}}(x, T;\e)}+\widetilde N_2\right),\label{N2psi}
\end{align}
where $\widetilde N_1$ and~$\widetilde N_2$ are given by
\begin{align}
& \widetilde N_1=\left(e^{-\frac12\sum_{k\geq1}\frac{\zeta^{k m}-\xi^{k m}}{k m z^{k m}}}-1\right)\frac{X_1(\xi,\zeta)\tau_{\mathrm{s}}(x,T;\e)}{\tau_{\mathrm{s}}(x,T;\e)},\nn\\
& \widetilde N_2=\left(e^{-\frac12\sum_{k\geq1}\frac{\zeta^{k n}-\xi^{kn}}{k m z^{k m}}}-1\right)
\frac{\zeta}{\xi}\frac{X_2(\xi,\zeta)\tau_{\mathrm{s}}(x, T;\e)}{\tau_{\mathrm{s}}(x, T;\e)}. \nn
\end{align}
Now by using \eqref{Yk-N1-N2} together with \eqref{Lpsi-tau}, \eqref{N1psi} and \eqref{N2psi} 
we arrive at
\begin{align}
- \frac{\left(Y_k\right)_-\psi_1}{\psi_1}
 =&\left(e^{-\sum_{\mu\in\I_1}\frac {z^{-\mu m}}{\mu m}\e\frac{\p}{\p T_\mu}}-1\right) 
 \Biggl[\frac{1}{\tau_{\mathrm s}(x,T;\e)}\biggl(\frac1{2m}\res_{\xi}\xi^{km+1}\p_\zeta^2 X_1(\xi,\zeta)\Big|_{\zeta=\xi}  \nn\\
& +\frac1{2n}\res_\xi \xi^{kn+1}\p_\zeta^2\left(\frac{\zeta}{\xi}X_2(\xi,\zeta)\biggr)\Big|_{\zeta=\xi} 
-\frac{\Gamma(m)\Gamma(n)}{\Gamma(1+h)}\frac{\p}{\p T_{k+1}}\right)
\tau_{\mathrm s}(x,T;\e)\Biggr], \nn
\end{align}
which leads to the formula~\eqref{Ypsi1} via a straightforward calculation. 
The proof of the formula~\eqref{Ypsi2} is similar. The lemma is proved.
\end{prf}

We are ready to prove the following theorem on the Virasoro constraints of the topological tau-function $\tau_{\rm top}$. 
In the proof of the theorem the notations $\psi_1,\psi_2, Y_k (k\geq 0)$, etc. 
correspond to~$u_{\rm top}$.
\begin{thm}\label{vira-cons-fvh}
The topological tau-function $\tau_{\rm top}$ satisfies the following Virasoro constraints:
\beq
L_k\tau_{\rm top}=0,\quad k\geq0.
\eeq
\end{thm}
\begin{prf}
Recall from Proposition~\ref{equivdzs} that 
\[
\tau_{\rm top}(x,T;\e)=\tau_{\mathrm s}\left(x+\frac{m\e}{2},T;\e\right).
\]
From Theorem~\ref{existencetautopo} we know
that $L_0\tau_{\rm top}=0$, which is, by a shift $x\mapsto x-\frac{m\e}{2}$, equivalent to 
\[
\left(L_0-\frac{x}{2nh\e}+\frac{m}{8nh}\right)\tau_{\mathrm s}=0.
\]
Then it follows from Lemma~\ref{lem-yl} that 
\beq\label{Y0psi}
-\left(Y_0\right)_-\psi_1=0,\quad \left(Y_0\right)_+\psi_2=-\frac{1}{2m}\psi_2.
\eeq
Since $-\left(Y_0\right)_-$ and $\left(Y_0\right)_+$ can be represented
in the form
\[
-\left(Y_0\right)_-=\sum_{k\leq-1}\gamma_k L^{k h/m},\quad
\left(Y_0\right)_+=\sum_{k\geq0}\gamma_k L^{-k h/n},
\]
where $\gamma_k \in \CC((\e))[[x-1, T]]$,  $k\in\ZZ$, 
the formulae in \eqref{Y0psi} imply that 
$\gamma_k=-\frac{1}{2m}\delta_{k,0}$, which yields
$Y_0=-\frac{1}{2m}$. 
So it follows from~\eqref{Ym-def} that 
\beq\label{aboveformulaeee}
Y_k=Y_0 L^{k h}=-\frac{1}{2m}L^{kh},\quad \forall \, k\geq1. 
\eeq
By using~\eqref{aboveformulaeee} together with Lemma~\ref{lem-yl} and
the identities
\begin{align*}
&\frac{-\left(L^{k h}\right)_-\psi_1}{\psi_1}
=\left(e^{\sum_{\mu\in\I_1}\frac {z^{-\mu m}}{\mu m}\e\frac{\p}{\p T_\mu}}-1\right)
\frac{\e\p_{T_k}\tau_{\mathrm s}}{\tau_{\mathrm s}},\\
&\frac{\left(L^{k h}\right)_+\psi_2}{\psi_2}
=\left(e^{h\e\p_x}e^{\sum_{\mu\in\I_2}\frac {z^{-\mu n}}{\mu n}\e\frac{\p}{\p T_\mu}}-1\right)
\frac{\e\p_{T_k}\tau_{\mathrm s}}{\tau_{\mathrm s}},
\end{align*}
we obtain for arbitrary $k\geq1$ that 
\begin{align}
&\left(e^{\sum_{\mu\in\I_1}\frac {z^{-\mu m}}{\mu m}\e\frac{\p}{\p T_\mu}}-1\right)
\frac{\left(\widetilde L_k+\frac{\e}{2m}\frac{\p}{\p T_k}\right)\tau_{\mathrm s}}{\tau_{\mathrm s}}=0, \nn \\
&\left(e^{h\e\p_x}e^{\sum_{\mu\in\I_2}\frac {z^{-\mu n}}{\mu n}\e\frac{\p}{\p T_\mu}}-1\right)\frac{\left(\widetilde L_k+\frac{\e}{2m}\frac{\p}{\p T_k}\right)\tau_{\mathrm s}}{\tau_{\mathrm s}}=0. \nn
\end{align}
This implies
\[
\left(L_k-\frac{\e}{2n}\frac{\p}{\p T_k}\right)\tau_{\mathrm s}=C_k(\e) \, \tau_{\rm s},
\quad C_k(\e)\in\CC[[\e]],
\]
which, after making the shift $x\mapsto x+\frac{m\e}2$, is equivalent to
\[
L_k\tau_{\rm top}=C_k(\e)\tau_{\rm top}.
\]
From the commutation relations
\[
\left[L_0,L_k\right]=-k L_k,
\]
it follows that $C_k(\e)=0$ for any $k\geq1$. 
The theorem is proved.
\end{prf}

\section{Proof of Theorem~\ref{main-1}}\label{Proofofmain}
Let us first prove Theorem~\ref{main-1} for the case 
$p=1/m,q=1/n,r=-1/h$, where $m,n$ are coprime positive integers and $h=m+n$.
To this end we will prove the following proposition, which generalizes part of the results of~\cite{DLYZ-2,DuY}.
\begin{prp} \label{main-2}
The following identity holds true:
\beq\label{mainidentity}
 Z(x, T; \e) = \tp(x, T; \e),
\eeq
where $Z(x, T; \e)$ and $\tp(x, T; \e)$ are defined in \eqref{defZ} and Theorem \ref{existencetautopo} respectively.
\end{prp}
\begin{prf}
By using the genus expansions~\eqref{genusexpansionZH}, \eqref{zh-15} 
and comparing the coefficients of~$\e^{2g-2}$ of the logarithms of both sides of~\eqref{mainidentity}, 
we find that the identity~\eqref{mainidentity} 
is equivalent to the following infinitely many identities:
\beq\label{fhg01234}
\F_g(x,T) = \widehat\F_g(x, T), \quad g\geq 0,
\eeq
where $\widehat\F_g(x, T)$ are defined by
\beq\label{defwcF}
\widehat\F_g(x, T) := (mnh)^{g-1} \mathcal H_g\bigl(t(x, T);1/m,1/n,-1/h\bigr) + \delta_{g,0} \frac{A}{mnh}.
\eeq

Let us first prove the identity \eqref{fhg01234} for $g=0$ by using the procedure that is similar to the one used in \cite{DuY}. 
Indeed, it is well known that $\H_0(t;1/m,1/n,-1/h)$ has the following explicit expression \cite{Du1,DLYZ-1}:
\beq\label{h-0}
\H_0(t;1/m,1/n,-1/h)=
\frac{1}{2}\sum_{i,j\geq0}\tilde t_i \tilde t_j\frac{\vh(t)^{i+j+1}}{(i+j+1)i!j!},\quad \tilde{t}_i=t_i-\delta_{i,1},
\eeq
where $\vh(t)\in \CC[[t]]$ is determined by the following equation:
\beq\label{kdvgenus0el}
\vh(t)=\sum_{i\geq0}t_i\frac{\vh(t)^i}{i!}. 
\eeq
Under the time substitutions~\eqref{time-trans} we find that 
$\vh(t(x,T))$ satisfies the equation
\beq
\sum_{\lambda\in\I}\lambda c_\lambda \widetilde T_\lambda e^{\lambda \vh(t(x,T))}+\frac{x}{mn}=0.
\eeq
By comparing this equation with~\eqref{el-equation-0} and by noticing that the solution to~\eqref{el-equation-0} in $\CC[[x-1,T]]$
must be unique we arrive at the following simple but crucial identity:
\beq\label{wvequal}
v_{\rm top}(x,T)=\frac1{m}\vh( t(x,T)),
\eeq
where we recall that $v_{\rm top}(x,T)=:v_{\rm top}$ is the topological solution to the dispersionless FVH.
Substituting \eqref{time-trans} and \eqref{wvequal} into \eqref{h-0}, and using~\eqref{def-f-0}, 
we find, after a lengthy but straightforward 
calculation that is similar to the one given in \cite{DuY}, that 
\beq\label{zh-17}
\F_0(x, T)=\widehat\F_0(x, T).
\eeq

Next we are to show that the power series $\F_g$ and $\widehat\F_g$ with $g\geq 1$ satisfy the 
technical conditions required by the uniqueness theorem proved in~\cite{LYZZ}. 
For $\F_g$, this has been given in the statement 
of Theorem~\ref{existencetautopo}: there exist functions 
$F_g(z_0,\dots,z_{3g-2})$, $g\geq1$ satisfying \eqref{tech1} and \eqref{tech2} such that   
\beq\label{F1F1again}
\F_g(x,T) =F_g\Bigl(v_{\rm top}, v_{\rm top}', \dots, v_{\rm top}^{(3g-2)}\Bigr), \quad g\geq 1.
\eeq
For~$\widehat \F_g$, the technical conditions were proved in \cite{DLYZ-1, DuY0, DZ-norm, LYZZ}: there exist functions  
\[\widehat F_g(z_0, z_1,\dots, z_{3g-2}), \quad g\geq 1\] 
such that 
\begin{align}
&\widehat \F_g (x,T) = \widehat F_g\left(m \vp(x, T), m \vp'(x, T),\dots, m \vp^{(3g-2)}(x, T)\right), \quad g\ge1.\label{tech3hodge}
\end{align}
Moreover, these functions can be chosen to satisfy the conditions
\begin{align}
& \widehat F_1= \widehat F_1(z_0,z_1) =  \frac{1}{24}\log z_1+\frac{\sigma_1}{24} z_0, \label{tech1hodge} \\
& \sum_{i\geq1} 
i z_i \frac{\p \widehat F_g}{\p z_i}=(2g-2) \widehat F_g, \quad \widehat F_g\in z_1^{-(4g-4)}\CC[z_1,\dots,z_{3g-2}], \quad g\geq 2.   \label{tech2hodge}
\end{align}
Here we use the relation~\eqref{wvequal} and the fact that $\p_x=\p_{t_0}$.

From Theorem~\ref{vira-cons-fvh} and the results of \cite{LYZZ} we know the validity of the following
Virasoro constraints: 
\begin{align}
\label{viratop} & L_k \tp=0,\\
\label{viraHodge} & L_k Z=0,
\end{align}
where $k\ge 0$, and the operators 
$L_k$ are defined in~\eqref{L0op} and~\eqref{zh-14}. 
According to \cite{LYZZ}, 
the equations 
\eqref{h-0}, \eqref{kdvgenus0el}, \eqref{tech3hodge} 
and \eqref{viraHodge}
imply that $\widehat{F}_g$, $g\ge 1$ satisfy the loop equation of \cite{LYZZ}. From \cite{LYZZ} we also know that the solution to the loop equation is unique under the 
conditions~\eqref{tech1hodge}, \eqref{tech2hodge}. 
On the other hand, by using the relations \eqref{wvequal},  \eqref{zh-17}, \eqref{tech3}, \eqref{el-equation-0} and \eqref{viratop} we know that $F_g$, $g\ge 1$ satisfy the same loop equation.
Then the fact that $F_g$ also satisfy the conditions \eqref{tech1} and \eqref{tech2} leads to the validity of the identity \eqref{mainidentity}. The proposition is proved.
\end{prf}

Proposition~\ref{main-2} implies the validity of Theorem~\ref{main-1} for the rational case.
Let us proceed to prove the theorem for the general case when $p,q,r\in\CC$ and satisfy
the local Calabi--Yau condition. To this end, let us first prove some lemmas.
\begin{lem}\label{lem-coef-1}
 The power series $u(x, T; \e) \in \CC[[x-1, T;\e]]$ defined by \eqref{def-u-cubic}
satisfies a hierarchy of evolutionary PDEs of the form
\beq\label{compare-cubic}
\frac{\p u}{\p T_\lambda}=\frac{\lambda}{pq} 
c_\lambda e^{\frac\lambda p u}
\left( u'+\sum_{g\geq1}\e^{2g}
\sum_{J\in\YY_{2g+1}}
C_{\lambda,J}(p,q) \, u^{(J)}\right),
\quad \lambda\in\I,
\eeq
where $u^{(J)}:=\p_x^{J_1} u\cdots \p_x^{J_{ l(J)}}u $, and $C_{\lambda,J}(p,q)\in\CC\bigl[p,q,p^{-1},q^{-1},(p+q)^{-1}\bigr]$, 
$J\in\YY$ are homogeneous rational functions of~$p$ and~$q$
of degree $1-|J|$, 
which are understood as the resulting expressions under the substitution $\lambda=kp$ 
or $\lambda=kq$, viewing the positive integer $k$ as a fixed parameter. 
\end{lem}
\begin{prf}
From~\cite{DLYZ-1} we know that
\beq\label{def-w-cubic926}
w:=-\frac{\e^2}{pqr} \p_{t_0}^2 \log Z_{\rm cubic}\left(t ;p,q,r;\frac{\sqrt{p+q}}{pq}\e\right)
\eeq
satisfies the Hodge hierarchy, which has the form
\[
\frac{\p w}{\p t_i}=X_i(w,w',\dots),\quad i\geq0
\] 
for some
$X_i(w,w',\dots)=:X_i\in\A_{w,0}[w',w'',\dots][[\e]]$. 
From \eqref{tech3hodge}--\eqref{tech2hodge} it follows that in the quasi-trivial transformation 
\[w=v+\Delta w=v+\e^2\p^2_{t_0} \widehat{F}_1(v, v')+\e^4\p^2_{t_0} \widehat{F}_2(v',v'',v^{(3)}, v^{(4)})+\dots\]
the function $\Delta w=\Delta w(v', v'',\dots)$
does not depend on $v$. So we know from the quasi-triviality of the Hodge hierarchy that the evolutionary equations for $w=w(t(x, T))$ have the forms
\beq\label{compare-cubic-w}
\frac{\p w}{\p T_\lambda}=\frac{\lambda}{pq} 
c_\lambda e^{\lambda w}
\left( w'+\sum_{g\geq1}\e^{2g}
\sum_{J\in\YY_{2g+1}}
\widetilde{C}_{\lambda,J}(p,q) \, w^{(J)}\right),
\quad \lambda\in\I,
\eeq
where the coefficients $\widetilde C_{\lambda,J}(p,q)\in\CC[p,q,p^{-1},q^{-1},(p+q)^{-1}]$.

Introduce a gradation on~$\CC[p,q,p^{-1},q^{-1},(p+q)^{-1}][[t]]$  via 
the following degree assignments:
\[
\overline{\rm deg}\, p=\overline{\rm deg}\, q=1,\quad \overline{\rm deg}\, t_i=i-1, ~ i\geq 0.
\]
Using the relations~\eqref{counting-dim}, 
we find that 
$\overline{\rm deg}\, \mathcal H_g(t;p,q,r)=3g-3$. 
Consequently, by using \eqref{time-trans} we obtain $\overline{\rm deg}\,w=-1$ as well as 
\[
\overline{\rm deg}\,\widetilde C_{\lambda,J}(p,q)= l(J)-|J|.
\] 
From~\eqref{def-u-cubic} and~\eqref{def-w-cubic926} we see that $u$ and~$w$ are related by a Miura-type transformation 
\beq\label{miura-trans}
u=p w+\frac{2p^2+2 p q+q^2}{24 p q^2} w'' \e^2+\cdots.
\eeq
Thus by using \eqref{compare-cubic-w} and \eqref{miura-trans} we obtain the
equations \eqref{compare-cubic} together with  
\[C_{\lambda,J}(p,q)\in\CC[p,q,p^{-1},q^{-1},(p+q)^{-1}].\]
Finally, from~\eqref{compare-cubic} and the fact that $\overline{\rm deg}\, u=0$,
we find that $C_{\lambda,J}(p,q)$ are homogeneous rational functions of $p, q$
satisfying 
$\overline{\rm deg}\,C_{\lambda,J}(p,q)=1-|J|$.
The lemma is proved. 
\end{prf}

\begin{lem}\label{lem-coef-2}
The flows of the FVH~\eqref{FVH-0} can be represented in the form
\beq
\frac{\p u}{\p T_\lambda}
=\frac{\lambda}{p q} c_\lambda e^{\frac\lambda p u}
\left(u'
+\sum_{g\geq1}\e^{2g} \sum_{J\in\YY_{2g+1}}
\overline C_{\lambda,J}(p,q) u^{(J)}\right),\quad \lambda\in \I,
\eeq
where $\overline C_{\lambda,J}(p,q)$ have the same properties as those of $C_{\lambda,J}(p,q)$ which appear in Lemma \ref{lem-coef-1}.
\end{lem}

\begin{prf}
One can prove this lemma by induction based on the following identity:
\[
\frac{\p}{\p T_{1/m}}\left(\frac{\p u}{\p T_\lambda}\right)
=\frac{\p}{\p T_\lambda}\left(\frac{\p u}{\p T_{1/m}}\right),
\quad \forall \lambda\in\I.
\]
We omit the details here. 
\end{prf}

For the reader's convenience, we give a few examples of $\overline C_{\lambda,J}(p,q)$ as follows:
\begin{align*}
& \overline C_{kp,(3)}(p,q)= \frac{k}{12} \frac{p+q}{p q^2}, \qquad
\overline C_{kp,(2,1)}(p,q)=\frac{k(2k+1)}{12} \frac{p+q}{p q^2},\\
& \overline C_{kq,(3)}(p,q)= \frac{k}{12} \frac{p+q}{p^2 q}, \qquad
\overline C_{kq,(2,1)}(p,q)=\frac{k}{12} \frac{(p+q)(2kq+p)}{p^3 q}.
\end{align*}

Now by using the validity of Theorem~\ref{main-1} 
for the rational case and by using Lemmas \ref{lem-coef-1}, \ref{lem-coef-2} we know that the equalities
\[
C_{\lambda,J}\biggl(\frac1m,\frac1n\biggr)
=\overline C_{\lambda,J} \biggl(\frac1m,\frac1n \biggr),\quad \forall \,J\in\YY
\]
hold true for arbitrary coprime positive integers $m$ and $n$. 
Then from the homogeneity of $C_{\lambda,J}(p,q)$ and $\overline C_{\lambda,J}(p,q)$ it follows the
validity of the identities 
\beq
C_{\lambda,J}(p,q)
=\overline C_{\lambda,J}(p,q),\quad \forall \, J\in\YY
\eeq
for arbitrary non-zero $p, q\in\CC$.
Hence the power series~$u(x,T;\e)$ given by the formula~\eqref{def-u-cubic}
is a solution to the FVH~\eqref{FVH-0}. 

To complete the proof of Theorem \ref{main-1}, we need to show that
the power series $Z(x, T;\e)$ given by the formula \eqref{defZ} 
is a tau-function of the FVH, i.e. to verify that it satisfies  
the relations \eqref{taudef2} and \eqref{taudef3}. 
Let us introduce, for any given $\lambda,\mu\in\I$, 
the following power series in $\CC[[x-1,T;\e]]$:
\begin{align*}
&\Delta_\lambda:=
\left(\lm_3-1\right)\lm_2^{-\frac12}
\e\frac{\p \log Z(x, T;\e)}{\p T_\lambda}
-\res_{\Lambda_3} \left.L^{-\lambda/r }\right|_{u=u(x, T;\e)},\\
&\Delta_{\lambda,\mu}:=
\e^2\frac{\p^2 \log Z(x, T; \e)}{\p T_\lambda\p T_\mu}
-\left.\Omega_{\lambda,\mu}\right|_{u=u(x, T;\e)},
\end{align*}
where $u(x, T;\e)$ is defined by \eqref{def-u-cubic}.
We are to show that
\beq\label{ns-5}
\Delta_\lambda=0, \quad \Delta_{\lambda,\mu}=0.
\eeq
Using the genus expansion of~$\log Z_{\rm cubic}$ given in~\cite{DLYZ-1} one can represent 
the two point correlation functions of~$Z_{\rm cubic}$ in the form
\beq\label{2-pt-hodge}
\frac{p+q}{p^2 q^2}\e^2\frac{\p^2 \log Z_{\rm cubic}}{\p t_i\p t_j}
\biggl(t(x,T);p,q,r; \frac{\sqrt{p+q}}{pq}\e \biggr)
=\frac{(v/p)^{i+j+1}}{(i+j+1)i!j!}+ X_{i,j}(v,v',\dots;\e)
\eeq
for $i, j\ge 0$. Here 
\[
v=v(x,T):=u(x,T;\e=0)\in\CC[[x-1,T]],
\] 
and $X_{i,j}(z_0,z_1,\dots;\e)=:X_{i,j}\in\e\A_{z_0,0}\left[z_1^{-1}, z_1,z_2,\dots\right][[\e]]$
satisfy the equation
\[
\sum_{k\geq1}k z_k\frac{\p X_{i,j}}{\p z_k}-\e\frac{\p X_{i,j}}{\p\e}=0.
\]
Then by using the definition~\eqref{defZ} of $Z(x, T;\e)$ and the formulae \eqref{2-pt-hodge} we arrive at 
\begin{align*}
\e^2\frac{\p^2\log Z(x, T;\e)}{\p x\p T_\lambda}
&=-r c_\lambda+\sum_{i\geq0}\frac{\p t_i}{\p T_\lambda}\e^2\frac{\p^2\log Z_{\rm cubic}}{\p t_0\p t_i}
\left(t(x, T);p,q,r;\frac{\sqrt{p+q}}{pq}\e\right)\\
&=-rc_\lambda e^{\lambda v/p}+ f_\lambda(v,v',\dots;\e),\\
\e^2\frac{\p^2\log Z(x, T;\e)}{\p T_\lambda\p T_\mu}
&=-\frac{r}{pq}\frac{\lambda \mu}{\lambda+\mu}
+\sum_{i,j\geq0}\frac{\p t_i}{\p T_\lambda}\frac{\p t_j}{\p T_\mu}
\e^2\frac{\p^2\log Z_{\rm cubic}}{\p t_i \p t_j}\left( t(x, T);p,q,r;\frac{\sqrt{p+q}}{pq}\e\right)\\
&=\frac1{p+q}\frac{\lambda\mu}{\lambda+\mu} c_\lambda c_\mu 
e^{(\lambda+\mu)v/p}+ f_{\lambda,\mu}(v,v',\dots;\e),
\end{align*}
where $f_\lambda(z_0,z_1,\dots;\e) =: f_\lambda$, $f_{\lambda,\mu}(z_0,z_1,\dots;\e)=:f_{\lambda,\mu}\in\e\A_{z_0,0}\bigl[z_1^{-1}, z_1,z_2,\dots\bigr][[\e]]$
satisfy the equations
\[
\sum_{k\geq1}k z_k\frac{\p f_\lambda}{\p z_k}-\e\frac{\p f_\lambda}{\p\e}=0,\quad
\sum_{k\geq1}k z_k\frac{\p f_{\lambda,\mu}}{\p z_k}-\e\frac{\p f_{\lambda,\mu}}{\p\e}=0.
\]
Hence by using Lemma~\ref{leading-omega}  
we find that there exist elements 
\beq\label{genuszerovanish}
\Delta_\lambda^*=\Delta^*_\lambda(z_0,z_1,\dots), ~
\Delta_{\lambda,\mu}^*=\Delta^*_{\lambda,\mu}(z_0,z_1,\dots) 
\in\e\A_{z_0,0}\bigl[z_1^{-1},z_1,z_2,\dots\bigr][[\e]] 
\eeq
satisfying the homogeneity conditions
\beq
\sum_{i\geq1}i z_i \frac{\p\Delta_\lambda^*}{\p z_i}-\e \frac{\p\Delta_\lambda^* }{\p\e}
=0,\quad \sum_{i\geq1}i z_i \frac{\p\Delta_{\lambda,\mu}^*}{\p z_i}-\e\frac{\p \Delta_{\lambda,\mu}^*}{\p\e}
=0,\label{delta-zero-22}
\eeq
such that
\[
\Delta_\lambda=\Delta^*_\lambda(v,v',\dots),\quad 
\Delta_{\lambda,\mu}=\Delta^*_{\lambda,\mu}(v,v',\dots).
\]
On the other hand, applying the operators $\e\p_{T_\lambda}$ and $\e^2\p_{T_\lambda}\p_{T_\mu}$ 
to both sides of~\eqref{def-u-cubic}, and using equations~\eqref{orproperty} and~\eqref{FVH-equiv}, we obtain
\[
\p_x \Delta_\lambda=0, \quad \p_x^2 \Delta_{\lambda,\mu}=0.
\]
Then by using the transcendency
of~$v(x,T)$ we have the identities (cf. Lemma~\ref{diff-poly-vani})
\beq\label{x-0}
\p \Delta^*_\lambda= 0, \quad  
\p^2\Delta^*_{\lambda,\mu}=0,
\eeq
where $\p=\sum_{i\geq0}z_{i+1}\frac{\p}{\p z_i}$.
Combining with \eqref{genuszerovanish}, \eqref{delta-zero-22}, we have 
$\Delta^*_\lambda= 0$ and $\Delta^*_{\lambda,\mu}=0$.
So the relations~\eqref{ns-5} hold true. 
Theorem~\ref{main-1} is proved. \hfill $\square$

\section{Proofs of Propositions~\ref{proppri}, \ref{p14} and Theorem~\ref{gapthm}}\label{GAP}
In this section, we first prove Proposition~\ref{proppri}. Then  
we do some explicit calculations for the coefficients~$R_g$ 
based on the Hodge--FVH correspondence, which leads to a proof of 
Proposition~\ref{p14}. 
Finally, we give a proof of Theorem~\ref{gapthm} 
for certain special cubic Hodge integrals.

\begin{prfn}{Proposition~\ref{proppri}}  
In the rational case, we have $p=1/m$, 
$q=1/n$ for some coprime positive integers $m, n$.
By using~\eqref{stringtype} and~\eqref{dilaton} as well as Theorem~\ref{main-2}
 we find that $Z$ satisfies 
\[\sum_{\mu\in \I} (1-\mu)T_\mu \frac{\p Z}{\p T_\mu} 
+ x \frac{\p Z}{\p x} + \e \frac{\p Z}{\p \e}  - \frac{x^2}{2m n h\e^2}Z-\frac{\sigma_1}{24}Z
=0, \]
from which the properties \eqref{gap1}, \eqref{gap2} follow immediately.
 The degree estimate $\deg R_g(\sigma_1,\sigma_3)\leq 3g-3$ with~$g\geq 2$ 
 follows from Theorem 1.3 of \cite{DLYZ-1}.
To show~\eqref{Faber} we first note that for $g\geq 2$, 
\begin{align}
& \H_g(t;p,q,r) \label{hgpqr}\\
= & (-1)^{g-1} \sum_{n\geq 0} \sum_{i_1,\dots,i_n\geq 0} \frac{t_{i_1}\cdots t_{i_n}}{n!}
 \int_{\overline{\mathcal{M}}_{g,n}}  \psi_1^{i_1}\cdots\psi_n^{i_n} \lambda_g \lambda_{g-1} \lambda_{g-2}  \left( p^g q^{g-1}r^{g-2} + (p\leftrightarrow q \leftrightarrow r) \right) \nn\\
&  + (-1)^{g-1} \sum_{n\geq 0} \sum_{i_1,\dots,i_n\geq 0} \frac{t_{i_1}\cdots t_{i_n}}{n!}
 \int_{\overline{\mathcal{M}}_{g,n}}  \psi_1^{i_1}\cdots\psi_n^{i_n} \lambda_{g-1}^3 \, p^{g-1}q^{g-1}r^{g-1} + ~ \cdots ,  \nn
\end{align}
where 
\[(p\leftrightarrow q \leftrightarrow r):=p^g r^{g-1} q^{g-2}+q^g p^{g-1}r^{g-2}+q^g r^{g-1}p^{g-2}+r^g p^{g-1}q^{g-2}+r^g q^{g-1}p^{g-2},\] 
and the dots ``$\cdots$" denotes the terms in $\H_g(t;p,q,r)$, as polynomials of $p,q,r$, having degree strictly less than $3g-3$.
From \eqref{pqrmnh} and \eqref{sigma1sigma3mnh} it follows that the degrees defined for polynomials in~$p,q,r$ 
and the ones for polynomials in $\sigma_1$, $\sigma_3$ are consistent. 
As it was proved in~\cite{DLYZ-1}, the following formula holds true for $g\geq 2$:
\[
\sum_{n\geq 0} \sum_{i_1,\dots,i_n\geq 0} \frac{t_{i_1}\cdots t_{i_n}}{n!}
 \int_{\overline{\mathcal{M}}_{g,n}}  \psi_1^{i_1}\cdots\psi_n^{i_n} \lambda_g \lambda_{g-1} \lambda_{g-2} = 
 \frac{1}{2(2g-2)!} \frac{|B_{2g}|}{2g} \frac{|B_{2g-2}|}{2g-2} \frac{\p^{2g-2} v(t)}{\p t_0^{2g-2}},
\]
where $v(t):=\p_{t_0}^2 \mathcal H_0(t;p,q,r)$. 
Then the identity~\eqref{Faber} follows from Mumford's relation 
\[\lambda_{g-1}^3= 2\lambda_g\lambda_{g-1}\lambda_{g-2}\] (see~\cite{Mum}) 
and the fact that
\beq
\left.v(t(x,T))\right|_{T=0}=\log x.
\eeq
The proposition is proved.
\end{prfn}

We proceed to prove Proposition~\ref{p14}.

\begin{prfn}{Proposition~\ref{p14}}  
Recall that Lemma \ref{lem-initial-form} states that the difference equation~\eqref{eq-initial}, i.e. the equation
\beq
\sum_{0\leq\alpha_1\leq\cdots\leq\alpha_m\leq n} 
e^{\sum_{j=1}^{m} V\left(x+\alpha_j m\e- \left(j-\frac12\right) n\e;\e\right)} =
\binom{m+n}m x, \label{again}
\eeq
has a unique solution of the form
\beq\label{v-again}
V(x;\e)=\frac1m\log x+\sum_{g\geq1}\e^{2g} \frac{P_g(m, n)}{x^{2g}}.
\eeq
We have shown in~\cite{LYZZ} that
\beq\label{def-Rg}
\log Z_{\rm cubic}\biggl(t(x,0);\frac1m,\frac1n,-\frac1h;\e\biggr)=
\e^{-2}\H_0(t(x,0))+\frac{\sigma_1-1}{24}\log x+\sum_{g\geq2}\e^{2g-2}\frac{R_g}{x^{2g-2}}, 
\eeq
where $R_g=R_g(\sigma_1,\sigma_3)\in\QQ[\sigma_1,\sigma_3]$, $g\geq2$, and
$\sigma_1$, $\sigma_3$ are given in~\eqref{sigma1sigma3mnh}. From
 Theorem~\ref{main-1} it follows that
\[
u_{\rm top}(x,T;\e)=\biggl(\lm_3^{\frac12}-\lm_3^{-\frac12}\biggr) \, \biggl(\lm_1^{\frac12}-\lm_1^{-\frac12}\biggr)
\log Z_{\rm cubic} \biggl(t(x,T);\frac1m,\frac1n,-\frac1h; \e\sqrt{mnh} \biggr).
\]
So by using~\eqref{iniproblem} we obtain
\begin{align}
V(x;\e)=\left(\lm_3^{\frac12}-\lm_3^{-\frac12}\right)\left(\lm_1^{\frac12}-\lm_1^{-\frac12}\right)
\log Z_{\rm cubic}\biggl(t(x,0);\frac1m,\frac1n,-\frac1h; \e\sqrt{mnh}\biggr). \label{relation-Rg-Pg}
\end{align}
The proposition then follows from \eqref{v-again}--\eqref{relation-Rg-Pg}.
\end{prfn}

\begin{emp} Let us apply Proposition~\ref{p14} to compute $R_g$ for $g=2,3,4$.
We first derive the expressions of the differential 
polynomials $M_1^{[g]}=M_1^{[g]}(u',\dots,u^{(2g)})$ defined in \eqref{coef-of-fvh}, \eqref{M-expr} for $g=2,3,4$. 
By using~\eqref{tau-symmetry} we obtain
\begin{align}
M_1^{[1]}=&\frac{mn(2mh-n)}{24}u''
+\frac{m^3n(h+1)}{24}(u')^2, \nn \\
M_1^{[2]}=&\frac{mn\left(24m^2nh^2-4mh(2m^2+2mn+7n^2)+7n^3\right)}{5760}u^{(4)} \nn \\
&+\frac{m^3n(h+1)\left(12mnh-4m^2-6mn-9n^2\right)}{1440}u' u''' \nn\\
&+\frac{m^2n\left(36m^2nh^2-4mh(3m^2-2mn+8n^2)-12m^3-8m^2n-12mn^2+5n^3\right)}{5760}(u'')^2\nn\\
&+\frac{m^4n(h+1)\left(22mnh-(8m^2+13n^2)-4h\right)}{2880} (u')^2 u'' \nn\\
&+\frac{m^5 n(h+1)\left(5mnh-(2m^2-3mn+2n^2)-2h\right)}{5760}(u')^4. \nn 
\end{align}
Since the expressions for $M_1^{[3]}$ and $M_1^{[4]}$ are quite long, we omit them here.
By using \eqref{top-initial-genus0} and \eqref{Vform3} we obtain 
\begin{align*}
P_1=&\frac{nh(m-1)}{24m},\\
P_2=&-\frac{nh(m-1)}{960m}\left(4mnh+3m^2-3m n-3n^2-m\right),\\
P_3=&\frac{nh(m-1)}{72576m}\left(
8 m^2 n^2 h^2+4 m n h \left(44 m^2-17 m n-17 n^2\right)+95 m^4-75 m^3 n\right.\\
&\left.-48 m^2 n^2+54 m n^3+27 n^4 -m \left(49 m^2-13 m n-13 n^2\right)+m^2+m
\right).
\end{align*}
Here we also omit the long expression of~$P_4$.
Then it follows from Proposition~\ref{p14} that
\begin{align}
& R_2=-\frac{1}{1440}+\frac{13\sigma_1}{5760}-\frac{7\sigma_1^2}{5760}
+\frac{1}{17280}\left(\sigma_1^3-\frac{\sigma_3}{2}\right), \nn \\
& R_3=
\frac{1}{181440}-\frac{107 \sigma_1}{362880}+\frac{145 \sigma_1^2}{290304}
-\frac{31 \sigma_1^3}{161280} \nn \\
&\qquad -\left(\frac{31}{1088640}-\frac{113 \sigma_1}{4354560}\right)\left(\sigma_1^3-\frac{\sigma_3}{2}\right)
-\frac{\left(\sigma_1^3-\sigma_3/2\right)^2}{13063680}, \nn
\end{align}
which coincide with the ones given in~\cite{LYZZ}.
In the same way we also obtain the following expression of $R_4$:
\begin{align}
& R_4=
\frac{211}{10886400}+\frac{\sigma_1}{48600}-\frac{1193 \sigma_1^2}{4354560}
+\frac{18629 \sigma_1^3}{58060800}-\frac{127 \sigma_1^4}{1290240} \nn \\
&\qquad +\left(\frac{83}{3870720}-\frac{1657 \sigma_1}{32659200}+\frac{6469 \sigma_1^2}{261273600}\right) 
\left( \sigma_1^3-\frac{\sigma_3}2\right) \nn \\
&\qquad +\left(\frac{17}{65318400}+\frac{247 \sigma_1}{1567641600}\right) \left( \sigma_1^3-\frac{\sigma_3}2\right)^2
-\frac{\left(\sigma_1^3-\sigma_3/2\right)^3}{2351462400}. \nn
\end{align}
\end{emp}

Let us now consider the case when one of $p,q$ is taken to be~1 and give a proof of Theorem~\ref{gapthm}.
It suffices to consider the case with $p=1$. So in what follows we assume that
\beq
(p,q,r)=\left(1,q,-\frac{q}{q+1}\right).
\eeq
We note that in this case the parameter $1/q$ is identified with the framing
parameter~$f$. For more details about the framing, see for example~\cite{Bri}. 

We need the following lemma.

\begin{lem}\label{lemmavlog}
For the rational case with $p=1$, the initial value $V(x;\e)$ for $u_{\rm top}(x,T;\e)$ has the following explicit expression:
\beq\label{Vlogx}
V(x;\e)\equiv \log x.
\eeq
\end{lem}
\begin{prf}
In the present case, the equation~\eqref{eq-initial} reads
\[
\sum_{i=0}^n \Lambda^{i-n/2} e^{V}=(n+1)x,
\]
whose unique solution in $\CC[[x-1;\e]]$ is clearly given by $V=\log x$.
The lemma is proved.
\end{prf}

By employing Lemma~\ref{lemmavlog} and Proposition~\ref{p14} 
one can obtain the following explicit expressions for the 
polynomials $R_g$:
\begin{align}
&R_1=-\frac1{24}\left((q+1)+(q+1)^{-1}\right), \label{idr1}\\
&R_g= -(2g-3)!\left(\widetilde B_g^2 +\sum_{i=1}^g \widetilde B_{g-i} \widetilde B_{g+i}\left((q+1)^i+(q+1)^{-i}\right)\right), 
\quad g\geq 2,\label{idrg}
\end{align}
where $\widetilde B_{i}:=\frac{1-2^{1-i}}{i!}B_{i}$, $i\geq0$.
By using \eqref{def-Hg}, \eqref{time-trans} and~\eqref{gap2} we know that $R_g$, $g\geq1$ are related to 
the cubic Hodge integrals by the formula 
\begin{align}
&R_g=\sum_{J\in\YY}
\frac{(-1)^{\ell(J)}}{m(J)!}
\int_{\M_{g,\ell(J)}}
\mathcal{C}_g(-1) \, \mathcal{C}_g(-q) \, \mathcal{C}_g\left(\frac{q}{1+q}\right)
\prod_{i=1}^{\ell(J)}\psi_i^{J_i+1} .
\end{align} 
Therefore, the formulae \eqref{idr1}, \eqref{idrg} yield relations for the Hodge integrals.  
These relations specialize to (1.18) of~\cite{DLYZ-2} when $q=1$.

Now let us prove Theorem~\ref{gapthm}.

\begin{prfn}{Theorem~\ref{gapthm}}  By using \eqref{defwcF} we know that it suffices to show that 
$(mnh)^{1-g}\widehat\F_g(x, T)$, $g\geq1$ possess the 
property given by~\eqref{stronggap1}, \eqref{stronggap2}.
According to Theorem~\ref{main-1}, $e^{\widehat\F(x,T;\e)}$ is the topological tau-function of the FVH. 
Due to the symmetry with respect to the 
interchange of $p,q$ in $\widehat\F_g(x, T)$, we only need to prove  
the theorem for the case $p=1$. 

Denote 
$\widehat\F(x, T_{\rm I};\e):=\widehat\F(x, T;\e)|_{T_{\rm II}=0}$, and 
introduce the $k$-point correlators for $\widehat\F(x,T_{\rm I};\e)$ as
follows:
\beq
C_{i_1,\dots,i_k}(x,\e) := \e^k \frac{\p^k \widehat\F(x,T_{\rm I};\e)}{\p T_{i_1}\dots\p T_{i_k}} \bigg|_{T_{\rm I}=0}, \quad k\geq1, \, i_1,\dots,i_k\geq1.
\eeq
From Proposition~\ref{proppri} we see that in order to prove the theorem,
we only need to show that all these correlators belong to 
$\CC[[x,\e]]$.
To this end, let us first look at the two-point correlators. 
Using the relation~\eqref{taudef3} between the tau-function of the 
FVH and two-point correlation functions
$\Omega_{\lambda,\mu}$, we obtain
\[
C_{i_1,i_2}(x,\e) =\Omega_{i_1,i_2}|_{T=0}, \quad i_1,i_2\geq 1.
\]
From the definition~\eqref{defomega} with $p=1$ 
it follows that $\Omega_{i,j}$ are polynomials of 
$\Lambda^{\frac12+\ell} e^{u_{\rm top}}$, $\ell\in \ZZ$.
Then by using Lemma~\ref{lemmavlog} we arrive at the fact 
that the 2-point correlators $C_{i_1,i_2}(x,\e) \in \CC[[x,\e]]$. 
Now let us introduce the notations
\[\Omega_{\lambda_1,\dots,\lambda_k} = \e^{k-2}\p_{T_{\lambda_1}} \dots \p_{T_{\lambda_{k-2}}} \Omega_{\lambda_{k-1},\lambda_k}, \quad 
k\geq 3. \] 
By using~\eqref{FVH-equiv} and by induction on~$k$ 
we know that $\Omega_{i_1,\dots,i_k}$ with 
$k\geq 2$ are all polynomials of 
$\Lambda^{\frac12+\ell} e^{u_{\rm top}}$, $\ell\in \ZZ$.
So from Lemma~\ref{lemmavlog} it follows
that $C_{i_1,\dots,i_k}(x,\e) \in \CC[[x,\e]]$ for all $k\geq 3$.
It remains to show that  the 1-point correlators also 
belong to $\CC[[x,\e]]$. To this end, let us use the  equation~\eqref{stringtype}. Dividing this equation 
by $Z=\tau_{\rm top}$, taking its derivative with respect to $T_{i}$, and then putting $T=0$, we find that 
$C_{i}(x,\e) \in \CC[[x,\e]]$, $i\geq1$.
The theorem is proved.
\end{prfn}

\begin{center}
{\small
Si-Qi Liu, Department of Mathematical Sciences, Tsinghua University, Beijing 100084, P.R.~China\\ 
e-mail: liusq@tsinghua.edu.cn\\
~\\
Di Yang, School of Mathematical Sciences, University of Science and Technology of China,\\
Hefei 230026, P.R.~China\\
e-mail: diyang@ustc.edu.cn\\
~\\
Youjin Zhang, Department of Mathematical Sciences, Tsinghua University, Beijing 100084, P.R.~China\\
e-mail: youjin@tsinghua.edu.cn\\
~\\
Chunhui Zhou, School of Mathematical Sciences, University of Science and Technology of China,\\ 
Hefei 230026, P.R.~China\\
e-mail: zhouch@ustc.edu.cn}
\end{center}

\end{document}